\documentclass[preprint,floats,tightenlines,11pt,eqsecnum,showpacs,aps]{revtex4}
\usepackage{amsmath}
\usepackage{graphicx}
\begin{document}
\def\appls{\hbox{$<$\kern-.75em\lower 1.00ex\hbox{$\sim$}}}






\title{QUANTUM STRUCTURE OF SPACETIME AND ITS ENTROPY\\ IN A CYCLIC UNIVERSE WITH NEGATIVE CURVATURE I:\\
A THEORETICAL FRAMEWORK}


\author{Miloslav Svec\footnote{svec@hep.physics.mcgill.ca}}
\affiliation{Physics Department, Dawson College, Westmount, Quebec, Canada H3Z 1A4}

\date{November 20, 2018}

\begin{abstract}

We construct a model of the Cyclic Universe from a joint theory  of General relativity, Thermodynamics and the Quantum information theory. Friedmann equations and the thermodynamical Gibbs-Duhem relation determine a general form of the Hubble function which predicts a dynamical Dark Energy (DE) and a dynamical Dark Matter (DM) described by new entropic terms and by the equations of state $w_0=-1$ and $w_M=0$, respectively, at all $z$. The entropic terms give rise to the acceleration and deceleration stages of the expansion of the Cyclic Universe.

We posit the spacetime has a quantum structure described by the Quantum information theory. We identify the space quanta $\rho$ with two-qubit quantum states of massless gravitons with helicity states $|\pm2>$. All space quanta carry quantum information entropy $S(\rho)$. All entangled quanta carry entanglement entropy $S_E(\rho)$ and form DE. All non-entangled quanta form DM. The average quantum state of DE is a special state  $\rho_\lambda(t)$. It is described by the scale factor $a(t)$ and carries entropy $\Sigma_\lambda(t)$. In the absence of Baryonic matter DM and DE are described by probability distributions of their entropies $p(\vec{x},t,S)$ and $q(\vec{x},t,\chi)$ where $\chi(\rho)=S_E(\rho)+S(\rho)$. Fisher information metric generates from these distributions the vacuum gravitational fields  $h^{MV}_{\mu\nu}$ and $h^{EV}_{\mu\nu}$ of DM and DE. In the presence of the Baryonic matter the distributions are displaced $p\to p'$ and $q\to q'$. Fisher metric then defines the displaced fields $h^{MB}_{\mu\nu}$ and $h^{EB}_{\mu\nu}$. In Einstein's theory of General relativity Space is the gravitational field which we identify with Dark Energy and Dark Matter fields. The theory predicts the existence of a new "residual" matter term with equation of state $w_r=-\frac{1}{3}$ in the Hubble function and the negative spatial curvature $k=-1$ the consequence of which are constraints on cosmological parameters. The theory also relates the new entropic terms of DE and DM in the Hubble function to the entropy $\Sigma_\lambda(t)$. We derive equations of state of Dark Energy and Dark Matter and the "residual" matter term from the kinetics of the space quanta modeled as non-classical particles with momentum and energy defined in terms of their entropies. We recover Robertson-Walker metric and the Friedmann equations from the gravitational fields of Dark Energy and Dark Matter. The predictions of the theory are tested and confirmed by cosmological data in Part II of this work.

\end{abstract}
\pacs{9880.-k, 9880.Qc, 9535.+d, 9535.+x}

\maketitle

\tableofcontents

\section{Introduction}

The key insight of Albert Einstein in his 1916's epoch making paper~\cite{einstein16} was the realization that space is not a "container" of the gravitational field: space {\it is} the gravitational field. This means that space is not something separate from the matter but one of the "material" components of the Universe. Einstein's equations of General Relativity relate these two Space/Gravitational Field and Matter components of the Universe.

Over the past century our view of the Universe has been evolving from seeing the Universe as a static and stable system to imagining it expanding at a constant velocity to assuming that this velocity is decreasing at a constant rate to recent observations that this velocity is actually accelerating~\cite{perlmutter97,perlmutter99,riess98,schmidt98} (for a recent review see~\cite{huterer17}). These observations are embodied in the Hubble function $H^2(z)$ of a highly successful $\Lambda$CDM Model~\cite{planck15} as a constant Dark Energy density term $\rho_\Lambda$ with equation of state $w_\Lambda=-1$ corresponding to a negative pressure $p_\Lambda=w_\Lambda\rho_\Lambda$. The physical origin of the Dark Energy is unknown in the $\Lambda$CDM Model and in the numerous alternative models of Dark Energy~\cite{amendola10,joyce16,brax18}. The observations to be made by the ongoing and upcoming astronomical surveys at high redshifts including Dark Energy Survey (DES)~\cite{DES}, Large Synoptic Survey Telescope (LSST)~\cite{LSST}, Euclid Mission~\cite{Euclid}, Wide Field Infrared Survey Telescope (WFIRST)~\cite{WFIRST,spergel15} and Square Kilometer Array (SKA)~\cite{SKA} will significantly contribute to our understanding of the Dark Energy and provide new tests of the $\Lambda$CDM Model~\cite{scott18}.

In the Part I of this work we construct a model of the Cyclic Universe from a joint theory of General relativity, Thermodynamics and the Quantum information theory of the Space based on four central propositions:\\
(1) Dark Energy and Dark Matter are the Space\\
(2) Space has an observable quantum structure\\
(3) Dark Energy and Dark Matter have distinct quantum structures\\
(4) Dark Energy and Dark Matter are gravitational fields arising from these quantum structures.\\
In the Part II of this work we test the predictions of this model by cosmological data~\cite{svec17b}. 
 
Friedmann equations and the thermodynamical Gibbs-Duhem relation determine a general form of the Hubble function which {\bf predicts} Dark Energy and Dark Matter terms in the Hubble function 
\begin{equation}
H^2 = H_0^2\Bigl[\Omega_{0,0}+\Sigma_0(z)
+(1+z)^3\Bigl(\Omega_{Mm,0}+\Sigma_M(z)\Bigr)
+(1+z)^4\Omega_{rad,0}\Bigr]\\
\end{equation}
The subscript $Mm$ indicates the inclusion of the atomic matter. The new entropic terms $\Sigma_0(z)$ and $\Sigma_M(z)$ are given by the entropies of Dark Energy $S_0(t)$ and Dark Matter $S_M(t)$, respectively. These entropic terms are related by the relation
\begin{equation}
\frac{d\Sigma_0}{dz}+(1+z)^3\frac{d\Sigma_M}{dz}
\end{equation}
which is equivalent to the entropy conservation $S_0(t)+S_M(t)=S'=const$. It is these entropic terms that give rise to the serial acceleration-deceleration transitions of the expansion of the Cyclic Universe in our Model A developed in Ref.~\cite{svec17a}. This General theory also determines the equations of state $w_0=-1$, $w_M=0$, $w_m=0$, $w_{rad}=+1/3$ at all z for Dark Energy, Dark Matter, atomic matter and radiation, respectively. The theory is self-consistent even when we add to the Hubble fuction an additional "residual" matter term with the equation of state $w_r=-\frac{1}{3}$ at all $z$
\begin{equation}
\tilde{H}^2=H^2 + H_0^2\Omega_{r,0}(1+z)^{3(1+w_r)}
\end{equation}

Since Space is a "material" substance it must have an observable quantum structure. The background spacetime is a standing gravitational wave $g_{\mu \nu}(\vec{x},t)$ given by the Roberson-Walker metric with a periodic scale factor $0<a_{min} \leq a(t)\leq a_{max}<\infty$ and tensor components $h_{ij}(\vec{x},t)=a(t)^2 S_{ij}(\vec{x})$. According to the principle of particle-wave duality this spacetime has a quantum structure. We identify the space quanta $\rho$ with two-qubit quantum states of massless gravitons with helicity states $|\pm2>$. All entangled space quanta form Dark Energy. All non-entangled space quanta form Dark Matter. All space quanta carry quantum information entropy~\cite{nielsen00,bengtsson06,vedral06} $S(\rho)$. All entangled quanta also carry relative entropy of entanglement~\cite{vedral06} $S_E(\rho)$ which defines the quantity $\chi(\rho)=S_E(\rho)+S(\rho)\geq0$. The entangled states of Dark Energy are non-local states that violate Bell inequality~\cite{bell64,bell87}.

In the absence of Baryonic matter the quanta of Dark Matter are described by the probability distribution $p(\vec{x},t,S)$ while the quanta of Dark Energy are described by the probability distribution $q(\vec{x},t,\chi)$. Fisher information metric~\cite{amari93} generates from these distributions vacuum gravitational fields of Dark Matter $h^{MV}_{\mu \nu}(\vec{x},t)$ and Dark Energy $h^{EV}_{\mu \nu}(\vec{x},t)$. In the presence of the Baryonic matter these distributions are displaced $p\to p'$ and $q\to q'$. Fisher metric then defines the displaced fields $h^{MB}_{\mu\nu}$ and $h^{EB}_{\mu\nu}$. Einstein's equations for these fields define the corresponding energy-momentum stress tensors. In general, on galactic and cluster scales we expect 
$|h^{EB}_{\mu \nu}| \ll |h^{MB}_{\mu \nu}|$. 

The fields $h^{MB}_{\mu\nu}$ and $h^{EB}_{\mu\nu}$ defined in a local inertial frame with Minkowski metric predict the existence of a new cosmological "residual" matter term $\rho_r(t)$ with equation of state $w_r=-\frac{1}{3}$. Consistency of this "internal" curvature term with the curvature in Robertson-Walker metric requires negative spatial curvature $k=-1$. The positivity of the spatial curvature density $\Omega_{c,0}>0$ imposes constraints on the cosmological parameters $\Omega_{0,0}$ and $\Omega_{Mm,0}$ in terms of parameters of $\Lambda$CDM Model
\begin{eqnarray}
\Omega_{Mm,0}+\Omega_{r,0} & > & \Omega_{Mm,0}^\Lambda\\
\Omega_{0,0} & < & \Omega_\Lambda
\end{eqnarray}
The average quantum state of the entangled states of Dark Energy is given by a special quantum state $\rho_\lambda(t)$ which is described by the scale factor $a(t)$. The entropy term $\Sigma_0(z)$ is a functional of the quantum information entropy $\Sigma_\lambda(z)$ of the state $\rho_\lambda(t)$. Together with the "residual" matter term, the joint dynamics of gravity, Thermodynamics and Quantum information theory enbodied in the entropic equations describes the Hubble function called Entropic Model E. 

In the Part II of this work we test the model E in fits to the Hubble data and angular diameter distance data~\cite{svec17b}. The values of $\chi^2/dof$ and  information criteria AIC and BIC are significantly better compared to the $\Lambda$CDM Model. The data support the quantum information model of Dark Energy and Dark Matter as the quantum structure of the spacetime. In a previous work~\cite{svec17a} we have developed an analytical model of a Cyclic Universe called Model A. The Model E is essentially identical to the Model A with an astonishing $\chi^2/dof=0.0000057$. We conclude that the Entropic Model E and the Model A represent the same Cyclic Universe with negative curvature.

The paper is organized as follows. Section II formulates the Cyclic Model of the Universe. Section III combines the Friedmann equations with Gibbs-Duhem relation to predict a general form of the Hubble function. In Section IV we apply the Laws of Thermodynamics to all components of the Universe separately. In Section V we formulate our model of Dark Energy and Dark Matter as the quantum structure of the spacetime. In Section VI we discuss the transformations of the entropies of Dark Energy and Dark Matter and calculate their entropic terms in a quantum model. In Section VII we derive equations of state of Dark Energy and Dark Matter and the "residual" matter term from the kinetics of the space quanta. In Section VIII we recover Robertson-Walker metric and in Section IX the Friedmann equations from the gravitational fields of Dark Energy and Dark Matter. The paper closes with our conclusions and outlook in Section X and two Appendices.

\newpage
\section{Friedmann equations of the Cyclic Universe}

The expanding and contracting homogeneous and isotropic spacetime of the Cyclic Universe is described by the Robertson-Walker (RW) metric. In cartesian coordinates it is given by~\cite{weinberg08,carroll04}
\begin{eqnarray}
g_{ij} & = & a^2(t)\delta_{ij}+a^2(t)\frac{k}{R_0^2}\frac{x^ix^j}{1-\frac{k}{R_0^2}\vec{x}^2}=a^2(t)\delta_{ij}+a^2(t)S_{ij}(\vec{x})\\
\nonumber
g_{00} & = & -1, g_{i0}=g_{0j}=0
\end{eqnarray}
where $R_0$ is the curvature parameter and $k=-1,0,+1$ stands for open, flat and closed geometry. It is a standig gravitational wave with tensor components $h_{ij}(\vec{x},t)=S_{ij}(\vec{x})a^2(t)$ and periodic non-singular scale factor $0<a_{min}\leq a(t) \leq a_{max} <\infty$. For a homogeneous and isotropic cosmic fluid with energy density $\rho$ and pressure $p$ Friedmann equations have the form
\begin{eqnarray}
\rho+\rho_\Lambda+\rho_c & = & \frac{3c^2}{8\pi G} H^2\\
p+p_\Lambda+p_c & = & \frac{3c^2}{8\pi G}\bigl(-H^2-\frac{2}{3}\frac{dH}{dt} \bigr)
\end{eqnarray}
Here $\rho_\Lambda$ and $p_\Lambda=-\rho_\Lambda$ are the energy density and pressure of cosmological constant and $\rho_c$ and 
$p_c=-\frac{1}{3}\rho_c$ are the energy density and pressure of the curvature~\cite{carroll04}. These two energy densities are given by
\begin{equation}
\rho_\Lambda = \frac{3c^2}{8\pi G} \frac{\Lambda}{3}, \quad
\rho_c = \frac{3c^2}{8\pi G}\frac{-kc^2}{R_0^2a^2}
\end{equation}
where $\Lambda$ is the cosmological constant. They satisfy continuity equations
\begin{eqnarray}
\frac{d\rho_\Lambda}{dt} +3H\rho_\Lambda & = & -3Hp_\Lambda\\
\nonumber
\frac{d\rho_c}{dt} +3H\rho_c & = & -3Hp_c
\end{eqnarray}
Using these relations Friedmann equations lead to similar continuity equation for the density $\rho$. The Hubble function is defined in terms of the scale factor
\begin{equation}
H(t)=\frac{1}{a(t)}\frac{da(t)}{dt}
\end{equation}
The cyclic scale factor is a complex wave function with a period $T$ so that $a(t+T)=a(t)$. During the expansion phase $H(t)>0$, during the contraction $H(t)<0$. At the turning points $t_\alpha=0$ and $t_\omega=T/2$ of the expanding  Universe the scale factor $a(t_\alpha)=a_{min}> 0$ and $a(t_\omega)=a_{max}< \infty$. Consequently
\begin{equation}
H(t_\alpha)=H(t_\omega)=0
\end{equation}
The contraction phase ends at the turning point $t_{2\alpha}=T$ with the scale factor $a(t_{2\alpha})=a_{min}$ and $H(t_{2\alpha})=0$. 

Since $H(t)$ is a cyclic function the combinations
\begin{eqnarray}
\bar{\rho}  & = & \rho + \rho_\Lambda + \rho_c\\
\nonumber
\bar{p} & = & p + p_\Lambda + p_c
\end{eqnarray}
are the cyclic energy density and the cyclic pressure. The Friedmann equations for the Cyclic Universe then read
\begin{eqnarray}
\bar{\rho} & = & \frac{3c^2}{8\pi G} H^2\\
\bar{p} & = & \frac{3c^2}{8\pi G}\bigl(-H^2-\frac{2}{3}\frac{dH}{dt} \bigr)
\end{eqnarray}
where $\bar{\rho}$ and $\bar{p}$ satisfy continuity equation 
\begin{equation}
\frac{d\bar{\rho}}{dt} +3H\bar{\rho} = -3H\bar{p}
\end{equation}
Notice that $H(t)$ does not depend on the curvature parameter $R_0$ and therefore on $\rho_c$, $p_c$.

The Hubble function determines the acceleration of the expansion of the Universe in terms of the deceleration parameter $q(t)$ or $q(z)$ given by
\begin{eqnarray}
q(t) & = & \frac{-1}{H^2a}\frac{d^2a}{dt^2}=-1-\frac{1}{H^2}\frac{dH}{dt}\\
q(z) & = &
-1+\frac{1+z}{H}\frac{dH}{dz}=-1+\frac{1}{2}\frac{1+z}{H^2}\frac{dH^2}{dz}
\end{eqnarray}

\section{Thermodynamics of the Universe}

\subsection{Friedmann equation for the Hubble function}

The Friedmann equations (2.9) and (2.10) determine the cosmic density $\bar{\rho}$ and pressure $\bar{p}$ for any given Hubble function. In this view the scale factor $a(t)$ and thus the Hubble function $H(t)$ are variables external to the theory of gravity. However, there is no physical theory of the scale factor that would supply the Hubble function.

We propose an alternate view in which the Hubble function is determined by the theory of gravity in terms of the pressure from the second Friedmann equation (2.10). The pressure in turn is determined by another fundamental theory: Thermodynamics. The cyclic energy density $\bar{\rho}$ is a derived variable from the first Friedmann equation (2.9). In this picture the evolution of the Universe is governed jointly by the Laws of gravity and the Laws of Thermodynamics.

The equation (2.10) can be written in the form
\begin{equation}
\bar{p} = \frac{3c^2}{8\pi G}\bigl(-H^2-\frac{a}{3}\frac{dH^2}{da} \bigr)
\end{equation}
or as a differential equation for $H^2$
\begin{equation}
\frac{dH^2}{da}+\frac{3}{a} H^2 =-\frac{8\pi G}{c^2} \frac{\bar{p}}{a}
\end{equation}
This is a first order linear equation $y'(x)+P(x)y=Q(x)$ for $y=H^2$ which has a solution~\cite{bronshtein15}
\begin{equation}
y(x)=\exp{\Bigl(- \int \limits_{x_0}^x P(x')dx'\Bigr)}
\Bigl[y(x_0) + \int \limits_{x_0}^x Q(x')
\exp{\Bigl(\int \limits_{x_0}^{x'} P(z)dz \Bigr)} dx'\Bigr]
\end{equation}
The solution of (3.2) then reads
\begin{equation}
H^2(a)=\Bigl(\frac{a_0}{a}\bigr)^3 \Biggl\{H_0^2-\frac{8\pi G}{c^2 a_0}
\int \limits_{a_0}^a \Bigl(\frac{a'}{a_0}\bigr)^2 \bar{p} da' \Bigg\}
\end{equation}
We shall refer to this equation as Friedmann equation for the Hubble function. It shows that the Hubble function is determined by the Laws of Thermodynamics.

\subsection{Gibbs-Duhem equation for the pressure}

We view the Universe as a thermodynamical system governed jointly by the Friedmann equations and the Laws of Thermodynamics. We thus supplement the Friedmann equations by the Euler's equation of the Thermodynamics~\cite{greiner94}
\begin{equation}
U=-\bar{p}V+kTS+\mu N +\sum \limits_i \mu_{r,i} N_{r,i}
\end{equation}
where $U$ is the internal energy of the Universe, $V$ is the expanding observable volume of the Universe, $T$ is its temperature, $S$ is its total entropy, $N$ its total number of particles and $N_{r,i}$ are additional (residual) extensive state variables. $k$ is the Boltzmann constant and $\mu$ is the chemical potential. The First Law of Thermodynamics requires that the expression $dU=-\bar{p}dV+kTdS+\mu dN +\sum \limits_i \mu_{r,i} dN_{r,i} $ be fully integrable. The total differential $dU$ then splits into two parts~\cite{greiner94}
\begin{eqnarray}
dU & = & -\bar{p}dV +kTdS + \mu dN +\sum \limits_i \mu_{r,i} dN_{r,i} \\
0  & = & -d\bar{p}V +kdT S + d\mu N + \sum \limits_i d\mu_{r,i} N_{r,i} 
\end{eqnarray}
The first equation is the First Law of Thermodynamics, the second equation is the Gibbs-Duhem relation~\cite{greiner94}. We assume that the Cyclic Universe is an isolated system in an equilibrium with $S=const$ and $N=const$ which satisfies the Second Law of Thermodynamics. Assuming at first that all $N_{r,i}=0$ the First Law reduces to the continuity equation (2.11). The Gibbs-Duhem relation then gives an independent expression for the pressure
\begin{equation}
\bar{p}(t)=\bar{p}_\alpha + \int \limits_{t_\alpha}^t \frac{kdT}{dt'}\frac{S}{V(t')}dt' +\int \limits_{t_\alpha}^t \frac{d\mu}{dt'}\frac{N}{V(t')}dt'
\end{equation}
With $\frac{kdT}{dt'}<0$ and $\frac{d\mu}{dt'}>0$  we expect periodic sign changing pressure reaching a maximum $\bar{p}_{max}>0$ at the point $\frac{d\bar{p}}{dt}=0$ where
\begin{equation}
\frac{kdT}{dt}\frac{S}{V} + \frac{d\mu}{dt}\frac{N}{V}=0
\end{equation}
In this model of cyclic $\bar{p}(t)$ we expect $\bar{p}_{min}=\bar{p}_\alpha$ and $\bar{p}_{max}=\bar{p}_\omega$. 

To evaluate the expression for $\bar{p}(t)$ we recall the well known relation for the temperature~\cite{ryden03,serjeant10}
\begin{equation}
a(t)T(t)=a(t_0)T(t_0)=a(t_\alpha)T(t_\alpha)
\end{equation} 
Taking a derivative we get
\begin{equation}
\frac{dT}{dt}=-\frac{a(t_\alpha)T(t_\alpha)}{a^2}\frac{da}{dt}
\end{equation}
The volume $V(t)=a^3(t)\mathcal{V}$ where $\mathcal{V}$ is the comoving volume of the Universe. We define comoving entropy density $\sigma=\frac{S}{\mathcal{V}}$ and comoving particle density $n=\frac{N}{\mathcal{V}}$. We express $\frac{d\mu}{dt}=\frac{d\mu}{da}\frac{da}{dt}$. With $\frac{da}{dt}dt=da$ we can write (3.7) as an integral over $a$
\begin{equation}
\bar{p}=\bar{p}_\alpha + \int \limits_{a_\alpha}^{a(t)} 
\Bigl[-kT(t_\alpha)a(t_\alpha)\frac{\sigma}{a^5} + \frac{d\mu}{da}\frac{n}{a^3} \Bigr] da
\end{equation}
To evaluate the integral over $\frac{d\mu}{da}$ we shall use the Generalized Mean Value Theorem~\cite{bronshtein15} that reads as follows: If the functions $f(x)$ and $\phi(x)$ are continous on the closed interval $[a,b]$, and $\phi(x)$ does not change sign in this interval, then there exists at least one point $\xi$, $a<\xi <b$, such that
\begin{equation}
\int \limits_a^b f(x)\phi(x)dx= f(\xi)\int \limits_a^b \phi(x)dx
\end{equation}
is valid. Thus we can write
\begin{equation}
\int \limits_{a_\alpha}^{a(t)} \frac{d\mu(a)}{da}\frac{n}{a^3} da=
\frac{d\mu(\xi(a))}{da}\int \limits_{a_\alpha}^{a(t)}\frac{n}{a^3} da
=-\frac{d\mu(\xi(a))}{da}\frac{n}{2}\Bigl(\frac{1}{a^2}-\frac{1}{a_0^2} \Bigr)
\end{equation}
where $\xi(a)\equiv\xi(a(t))$ indicates the dependence of the point $\xi$ on the variable upper limit of the integral. Carrying out the remaining integration we find
\begin{equation}
\bar{p}=\bar{p}(a)=\bar{p}_\alpha+
\frac{kT(t_\alpha)a(t_\alpha)\sigma}{4}\Bigl[\frac{1}{a^4}-\frac{1}{a^4_\alpha}\Bigr]  
-\frac{d\mu(\xi(a))}{da}\frac{n}{2}\Bigl[\frac{1}{a^2}-\frac{1}{a^2_\alpha}\Bigr]
\end{equation}
where $a\equiv a(t)$ is the time dependent scale factor. We are free to choose the lower limit of the integration and chossing the present time $t_0$ the equation (3.15) reads
\begin{equation}
\bar{p}=\bar{p}(a)=\bar{p}_0+
\frac{kT_0a_0\sigma}{4}\Bigl[\frac{1}{a^4}-\frac{1}{a^4_0}\Bigr]  
-\frac{d\mu(\xi(a))}{da}\frac{n}{2}\Bigl[\frac{1}{a^2}-\frac{1}{a^2_0}\Bigr]
\end{equation}

The key insight is the observation that logically the pressure is not determined by the Friedmann equations but originates in the internal thermodynamics of the Universe in which the entropy density and particle density of the Universe play important role. The equations (3.4) and (3.16) connect the two dynamical aspects of the evolution of the Universe: gravity and thermodynamics.

\subsection{General theory: The prediction of Dark Energy and Dark Matter}

With Gibbs-Duhem pressure (3.16) the Friedmann equation for the Hubble function (3.4) has an elegant and physically significant general solution. To carry out the integrations we shall use
\begin{eqnarray}
3\int \limits_{a_0}^a \bar{p}(a')a^{'2}da'
& = & \int \limits_{a_0}^a \frac{d}{da} \Bigl(\bar{p}(a')a^{'3}\Bigr)da'-
      \int \limits_{a_0}^a\frac{d\bar{p}}{da'}a^{'3}da'\\
\nonumber
& = & \Bigl(\bar{p}a^3-\bar{p_0}a_0^3\Bigr) 
-kT_0a_0 \sigma \Bigl(\frac{1}{a}-\frac{1}{a_0}\Bigr)-[\mu(a)-\mu(a_0)]n 
\end{eqnarray}
In terms of the redshift $\frac{a_0}{a}=1+z$ the general solution for the Hubble function then reads
\begin{eqnarray}
H^2 & = & H_0^2\Bigl[\Omega_{0,0}+\Sigma_0(z)
+(1+z)^3\Bigl(\Omega_{Mm,0}+\Sigma_{M}(z)\Bigr)+(1+z)^4\Omega_{rad,0}\Bigr]\\
\Omega_{rad,0} & = & \frac{8\pi G}{3c^2H_0^2}
\Bigl(\frac{3kT_0\sigma}{4a_0^3}\Bigr)
=\frac{8\pi G}{3c^2H_0^2}\Bigl(3p_{rad,0}\Bigr)
=\frac{8\pi G}{3c^2H_0^2}\rho_{rad,0}\\
\Omega_{0,0} & = & \frac{8\pi G}{3c^2H_0^2} \Bigl(-\bar{p}_0+\frac{kT_0\sigma}{4a_0^3}\Bigr)
=\frac{8\pi G}{3c^2H_0^2}\Bigl(-p_{0,0}\Bigr)
=\frac{8\pi G}{3c^2H_0^2}\rho_{0,0}\\
\Omega_{Mm,0} & = & \Omega_{M,0}+\Omega_{m,0}
=1-\Omega_{0,0}-\Omega_{rad,0}\\
\Sigma_0(z)  & = & \frac{8\pi G}{3c^2H_0^2}
\Bigl[\frac{d\mu(\xi(a))}{da} \frac{n}{2}\Bigr]\Bigl(\frac{1}{a^2}-\frac{1}{a_0^2}\Bigr)\\
\Sigma_{M}(z)  & = & \frac{8\pi G}{3c^2H_0^2}
\bigl[\mu(a)-\mu(a_0)\bigr]\frac{n}{a_0^3}
\end{eqnarray}
where $\bar{p}=p_0+p_{rad}$ is the total pressure (3.16) and $p_{0,0}$ and $p_{rad,0}$ are the present pressures of Dark Energy and radiation, respectively. The notation $\Omega_{Mm,0}$ indicates the inclusion of the atomic matter $m$. Notice that the correction terms $\Sigma_0(z)$ and $\Sigma_{M}(z)$ vanish at $z=0$. We refer to (3.18) and (3.19)-(3.23) as the General theory. An alternative derivation of (3.18) is given in the Appendix A.

The Hubble function (3.18) shows that the joint theory of gravity and thermodynamics naturally predicts the existence of Dark Energy $\tilde{\Omega}_0=\Omega_{0,0}+\Sigma_0(z)$ and Dark Matter
$\tilde{\Omega}_M=\Omega_{M,0}+\Sigma_M(z)$. Notably, the first term of the Dark Energy $\Omega_{0,0}$ is proportional to the present pressure. The second terms $\Sigma_0(z)$ and $\Sigma_M(z)$ describe the dependence of the Dark Energy and Dark Matter on the internal dynamics of the evolution in terms of changes of the chemical potential. Although the total number of the particles is constant, they undergo changes with the evolution of the Universe. Each component of the Universe is described by its own chemical potential $\mu_k(t)$, $k=0,M,m,rad$ such that
\begin{equation}
\frac{d\mu}{dt}N=\sum \limits_{k=0}^{rad} \frac{d\mu_k}{dt}N_k
\end{equation}
In the following we shall use the deceleration parameter (2.13) in the form 
\begin{equation}
q(z)=-1+\frac{1}{2}\frac{1+z}{H^2}\frac{dH^2}{dz}
\end{equation}
If in (3.18) the Dark Energy terms $\Omega_{0,0}$ and $\Sigma_0$ were zero and the Dark Matter term $\Sigma_M$ were zero, and the atomic matter term were $(1+z)^3\Omega_{m,0}$, then we would find 
\begin{equation}
q(z)=\frac{2\Omega_{M,0}+2\Omega_{m,0}+3\Omega_{rad,0}}{\Omega_{M,0}+\Omega_{m,0}+\Omega_{rad,0}} >1
\end{equation}
With $q(z)>0$ at all z there would be no accelerated expansion. Dark Energy terms are therefore necessary conditions for the acceleration to occur. A straightforward calculation shows that the sufficient condition $q(z=0)<0$ requires
\begin{equation}
2\Omega_{0,0}-\bigl(\frac{d\Sigma_0}{dz}+\frac{d\Sigma_M}{dz}\bigr)_{z=0} > \Omega_{M,0}+\Omega_{m,0}+2\Omega_{rad,0}
\end{equation}
which is clearly satisfied in the $\Lambda$CDM Model and in all models described in this work. The joint dynamics of gravity and thermodynamics thus naturally explains the decelerated and accelerated periods of the expansion of the Universe and confirms that the equation of state of the Dark Energy is $w_0=-1$ and that of  Dark Matter is $w_M=0$ at all $z$.

\subsection{The self-consistency of the General theory and its extension}

In the General theory the non-standard terms $\Sigma_M$ and $\Sigma_0$ of the Dark Matter and Dark Energy are not independent. Using the integral form (3.14) the expression (3.22) for Dark Energy reads
\begin{equation}
\Sigma_0(z)=-\frac{8\pi G}{3c^2H_0^2}\int \limits_{a_\alpha}^{a(t)} \frac{d\mu(a)}{da}\frac{n}{a^3} da
\end{equation}
Then from (3.23) follows the relation
\begin{equation}
\frac{d\Sigma_0}{da}=-\Bigl(\frac{a_0}{a}\Bigr)^3\frac{d\Sigma_M}{da}
\end{equation}
This important relation between Dark Matter and Dark Energy is central to the proof of the self-consistency of the General theory and its extension.

We extend the Hubble function (3.18) by a new term analogous to the atomic matter term but with a non-zero equation of state $w_r$
\begin{equation}
\tilde{H}^2=H^2+H_0^2\Omega_{r,0}(1+z)^{3(1+w_r)}
\end{equation}
To test the self-consistency of the General theory and its extension we consider two different forms of the total equation of state 
\begin{equation}
w=\frac{\bar{p}}{\bar{\rho}}=
\frac{-\rho_0+w_r\rho_r +\frac{1}{3}\rho_{rad}}{\bar{\rho}}   
\end{equation}
With deceleration parameter $q$ given by (3.25) they read
\begin{eqnarray}
w(1) & = & -\frac{1}{3}\Bigl(1-2q\Bigr)=
\frac{1}{3\tilde{H}^2}\Bigl\{-3\tilde{H}^2+(1+z)\frac{d\tilde{H}^2}{dz}\Bigr\}\\
w(2) & = & \frac{H_0^2}{\tilde{H}^2}\Bigl\{-(\Omega_{0,0}+\Sigma_0) 
+ w_r\Omega_{r,0}(1+z)^{3(1+w_r)} +\frac{1}{3}\Omega_{rad,0}(1+z)^4\Bigr\}
\end{eqnarray}
The theory is self-consistent when the ratio $W_0=\frac{w(1)}{w(2)}=1$. The calculation of $w(1)$ yields
\begin{equation}
w(1)=\frac{H_0^2}{3\tilde{H}^2}
\Bigl\{\frac{d\Sigma_0}{dz}+(1+z)^3\frac{d\Sigma_M}{dz}\Bigr\} + w(2)
\end{equation}
The relation (3.29) implies that $w(1)=w(2)$ so that the Extended General theory is self-consistent with $W_0=1$.

Finally we examine whether the Dark Energy and Dark Matter terms   $\tilde{\Omega}_0=\Omega_{0,0}+\Sigma_0$ and $\tilde{\Omega}_M=(1+z)^3\Bigl[\Omega_{M,0}+\Sigma_M\Bigr]$ in (3.18) can arise from a homogeneous and isotropic energy momentum stress tensor $T^{\mu \nu}$. The conservation of energy $T^{0 \nu}_{,\nu}=0$ implies~\cite{weinberg08}
\begin{equation}
\frac{d\rho}{dt}+3H(\rho+p)=0
\end{equation}
Integrating this equation using $dt=-\frac{dz}{(1+z)H}$ we find
\begin{equation}
\rho(z)=\rho_0\exp\Biggl[\int \limits_0^z \frac{3(1+w)}{1+z'}dz' \Bigr]
\end{equation}
where $w=w(z)=p(z)/\rho(z)$. For Dark Energy we set $w=w_0(z)=w_0(0)+\xi_0(z)=-1+\xi_0(z)$ and for Dark Matter $w=w_M(z)=w_M(0)+\xi_M(z)=\xi_M(z)$ to obtain
\begin{eqnarray}
\tilde{\Omega}_0(z) & = & \Omega_{0,0} + 
\Omega_{0,0}\Biggl\{\exp\Biggl[\int \limits_0^z \frac{3\xi_0}{1+z'}dz'\Biggr]-1\Biggr\}=\Omega_{0,0}+\Sigma_0\\
\tilde{\Omega}_M(z) & = & (1+z)^3\Biggl(\Omega_{M,0} + 
\Omega_{M,0}\Biggl\{\exp\Biggl[\int \limits_0^z \frac{3\xi_M}{1+z'}dz'\Biggr]-1\Biggr\}\Biggr)=(1+z)^3\Bigl[\Omega_{M,0}+\Sigma_M\Bigr]
\end{eqnarray}
The new terms $\Sigma_0$ and $\Sigma_M$ arise from the dependence of the equations of state $w_0(z)$ and $w_M(z)$ of Dark Energy and Dark Matter on the redshift $z$. To satisfy the condition (3.29) we require that $\xi_0=-(1+z)^3\xi_M$. However the terms $w(1)$ and $w(2)$ are now related by
$w(2)=w(1) + \xi_0\tilde{\Omega}_0(z)+\xi_M\tilde{\Omega}_M(z)$. The simple form of the homogeneous and isotropic $T^{\mu \nu}$ is consistent with the Friedmann equatiions only when $\xi_0=\xi_M=0$ which implies $\Sigma_0=\Sigma_M=0$, meaning that the chemical potential $\mu=const$.

To undestand this apparent inconsistency we recall that a process in which no heat is exchanged is called adiabatic process~\cite{greiner94}. It is described by the First Law $dU=-pdV$, i.e. by the same equation (3.27) as the conservation of the energy of the energy-momentum stress tensor. Since the entire Universe is an isolated system it both exchanges no heat and conserves its energy. So it is appropriate to describe this fact by treating the cosmic fluid as perfect fluid as well as an adiabatic system. Dark Energy and Dark Matter constantly exchange their Landauer heat of their quantum information entropy. This process is described by the non-adiabatic form of the First Law in the next Section. The adiabatic form of the energy conservation by $T^{\mu \nu}$ given by (3.27) does not need to apply to the interacting non-adiabatic components of the Universe just because it does apply to the non-interacting entire Universe.

\section{Entropic theory of the Hubble function.}

\subsection{Euler's equations for the four components of the Universe}

At first we shall assume there are four components of the Universe corresponding to Dark Energy, Dark Matter, atomic matter and radiation.  Originally Dark Matter and Dark Energy were discovered by the observations of the motions of the galaxies, stars and the supernova. The general solution (3.18) for the Hubble function naturally leads to the concepts of Dark Energy and Dark Matter theoretically represented by the terms $\Omega_{0,0}+\Sigma_0(z)$ and $(1+z)^3\Bigl(\Omega_{M,0}+\Sigma_{M}(z)\Bigr)$, respectively. Entropic theory is an extension of the Thermodynamics of the Universe which assumes that all components of Universe are thermodynamical systems governed by independent Euler's equations subject to the conservation of the total entropy $S$ and total particle number $N$. The Euler equation (3.5) can be written for each component $k=0,M,m,rad$
\begin{equation}
\rho_k+p_k=(1+w_k)\rho_k=\frac{kT}{V}S_k+\frac{\mu_k}{V}N_k
\end{equation}
where $w_k$ are equations of state. The First Law equations and the Gibbs-Duhem relations read
\begin{eqnarray}
\frac{d\rho_k}{dt}+3H\rho_k & = & -3Hp_k+\frac{kT}{V}\frac{dS_k}{dt}+
\frac{\mu_k(t)}{V}\frac{dN_k}{dt}\\
\frac{dp_k}{dt} & = & k\frac{dT}{dt}\frac{S_k(t)}{V} +
\frac{d\mu_k}{dt}\frac{N_k(t)}{V}
\end{eqnarray}
Adding these equations we find
\begin{eqnarray}
\frac{d\bar{\rho}}{dt}+3H\bar{\rho} & = & -3H\bar{p}+ 
\sum \limits_{k=0}^{rad} \Bigl\{\frac{kT}{V}\frac{dS_k}{dt} + 
\frac{\mu_k}{V}\frac{dN_k}{dt}\Bigr\}\\
\frac{d\bar{p}}{dt} & = & 
k\frac{dT}{dt}\sum \limits_{k=0}^{rad}\frac{S_k(t)}{V} + 
\sum \limits_0^{rad} \frac{d\mu_k}{dt}\frac{N_k(t)}{V}
\end{eqnarray}
The consistency of (4.4) with the continuity equation (2.11) requires 
\begin{equation}
\sum \limits_{k=0}^{rad} \Bigl\{\frac{kT}{V}\frac{dS_k}{dt} + 
\frac{\mu_k}{V}\frac{dN_k}{dt}\Bigr\}=\frac{kT}{V}\frac{dS}{dt} + 
\frac{\mu}{V}\frac{dN}{dt}=0
\end{equation}
The consistency of (4.5) with the Gibbs-Duhem relation (3.9) requires
\begin{equation}
\sum \limits_{k=0}^{rad} S_k(t)=S, \quad \sum \limits_{k=0}^{rad} \frac{d\mu_k}{dt}N_k(t)=\frac{d\mu}{dt}N
\end{equation}
where $N$ is the total number of particles in the Universe $N=\sum \limits_{k=0}^{rad}N_k(t)$. We assume that the Universe is an isolated system in an equilibrium with $S=const$ and $N=const$. The quantum information nature of the entropies $S_0$ and $S_M$ of Dark Energy and Dark Matter is discussed in the Section V.

\subsection{Special Entropic theory for the energy densities and the General theory}

The general solutions of the entropic equations (4.2) from the First Law are given by (3.3) and read
\begin{equation}
\rho_k(t) = \Bigl(\frac{a(t_0)}{a(t)}\Bigr)^{3(1+w_k)}
\Bigl[\rho_k(t_0)+\int \limits_{t_0}^t \Bigl(\frac{a(t')}{a(t_0)}\Bigr)^{3(1+w_k)}\Bigl\{\frac{kT}{V}\frac{dS_k}{dt} + 
\frac{\mu_k}{V}\frac{dN_k}{dt}\Bigr\} dt' \Bigr]
\end{equation}
where $w_k$ are again equations of state. The first terms in (4.8) are the Standard Model terms. The second terms in (4.8) are entropic terms given by the entropic integrals
\begin{equation}
I_k(t,t_0)=\frac{3c^2H_0^2}{8\pi G}\Sigma_k(t,t_0)=\int \limits_{t_0}^t \Bigl(\frac{a(t')}{a(t_0)}\Bigr)^{3(1+w_k)}\Bigl\{\frac{kT}{V}\frac{dS_k}{dt'} + \frac{\mu_k}{V}\frac{dN_k}{dt'}\Bigr\} dt'
\end{equation}
With normalized and unnormalized fractional energy densities $\Omega_k$ and $\tilde{\Omega}_k$ defined, respectively, by
\begin{equation}
\Omega_k=\frac{H_0^2}{H^2}\tilde{\Omega}_k=\frac{8 \pi G}{3 c^2 H^2} \rho_k
\end{equation}
we can write the equations (4.8) in the form
\begin{eqnarray}
\tilde{\Omega}_0(z) & = & (1+z)^{3(1+w_0)}\Bigl[\Omega_{0,0} + \Sigma_0(z)\Bigr]\\
\tilde{\Omega}_M(z) & = & (1+z)^{3(1+w_M)}\Bigl[\Omega_{M,0} + 
\Sigma_M(z)\Bigr]\\
\tilde{\Omega}_m(z) & = & (1+z)^{3(1+w_m)}\Bigl[\Omega_{m,0} + 
\Sigma_m(z)\Bigr]\\
\tilde{\Omega}_{rad}(z) & = & (1+z)^{3(1+w_{rad})}\Bigl[\Omega_{rad,0} + \Sigma_{rad}(z)\Bigr]\\
\end{eqnarray}
The Hubble function is given by
\begin{equation}
H^2(z)=H^2_0\sum \limits_{k=0}^{rad}\tilde{\Omega}_k(z)
\end{equation}
Next we identify the General theory with the Entropic theory and set 
\begin{eqnarray}
\Sigma_M & = & \frac{8\pi G}{3 c^2 H_0^2} (\mu(a)-\mu(a_0))\frac{n}{a_0^3}=\frac{8\pi G}{3 c^2 H_0^2} \int \limits_{a_0}^a \Bigl(\frac{a'}{a_0}\Bigr)^3 \frac{kT}{V}\frac{dS_M}{da}\\
\nonumber
\Sigma_0 & = & -\frac{8\pi G}{3 c^2 H_0^2}\int \limits_{a_0}^a \frac{d\mu(a)}{da}\frac{n}{a^3}da=\frac{8\pi G}{3 c^2 H_0^2}\int \limits_{a_0}^a  \frac{kT}{V}\frac{dS_0}{da}
\end{eqnarray}
From both relations we obtain
\begin{equation}
\frac{d\mu}{da}\frac{n}{a^3}= -\frac{kT}{V}\frac{dS_0}{da}=\frac{kT}{V}\frac{dS_M}{da}
\end{equation}
which implies $\frac{dS_0}{da}+\frac{dS_M}{da}=0$ or $S_0(t)+S_M(t)=S'=const$.
We can write the mass term in (3.18) in the form that separates Dark Matter and atomic matter
\begin{equation}
(1+z)^3\Bigl[\Omega_{Mm,0}+\Sigma_{Mm}(z)\Bigr]=
(1+z)^3\Bigl[\Omega_{M,0}+\Sigma_{M}(z)\Bigr]+
(1+z)^3\Bigl[\Omega_{m,0}+\Sigma_{m}(z)\Bigr]
\end{equation}
Comparison with the general solution (3.18) for the Hubble function shows that the equations of states predicted by the joint theory of gravity and Thermodynamics are $w_0=-1$, $w_M=w_m=0$ and $w_{rad}=\frac{1}{3}$ for $\it{all}$ $z$. 

In general entropic theory the solutions (4.1) and (4.8) for the energy densities $\rho_k$ depend on three unknown functions $S_k(t)$, $\mu_k(t)$ and $N_k(t)$. As a result the theory is not tractable and must be simplified using physically justified assumptions. 

(1) We shall assume that Dark Energy and Dark Matter have a constant total number of their particles (quanta) $N_0=const$, $N_M=const$. These particles are immutable: they do not transform into each other nor into the particles of the atomic matter or radiation. However these particles do change internally with the evolution of the Universe which leads to their non-zero chemical potentials $\mu_0(t)$ and $\mu_M(t)$. In contrast, atomic particles and particles of the radiation like photons and neutrinos do interact. As the result it is their sum that is constant $N_m(t)+N_{rad}(t)=N''=const$. 

(2) Next we shall assume that there are no transfers of the entropy between the dark and visible sectors. This means that $S_0(t)+S_M(t)=S'=const$ and $S_m(t)+S_{rad}(t)=S''=const$. 

(3) Assuming (1) and (2) it then follows from (4.6) that 
\begin{equation}
\Bigl(\frac{kT}{V}\frac{dS_m}{dt}+\frac{\mu_m}{V}\frac{dN_m}{dt}\Bigr)+
\Bigl(\frac{kT}{V}\frac{dS_{rad}}{dt}+\frac{\mu_{rad}}{V}\frac{dN_{rad}}{dt}\Bigr)=0
\end{equation}
Based on the comparison of (4.13) and (4.14) with the general solution (3.18) we assume $\Sigma_{m}=\Sigma_{rad}=0$ so that
\begin{equation}
\Sigma_m=\frac{kT}{V}\frac{dS_{m}}{dt} + \frac{\mu_m}{V}\frac{dN_{m}}{dt} =0
\end{equation}
\begin{equation}
\Sigma_{rad}=\frac{kT}{V}\frac{dS_{rad}}{dt} + \frac{\mu_{rad}}{V}\frac{dN_{rad}}{dt} =0
\end{equation}
With these three assumptions the final form of the entropic equations (4.2) in our fits reads
\begin{eqnarray}
\frac{d\rho_0}{dt} & = & \frac{kT}{V}\frac{dS_{0}}{dt}\\
\nonumber
\frac{d\rho_M}{dt} +3H\rho_M & = & \frac{kT}{V}\frac{dS_{M}}{dt}\\
\nonumber
\frac{d\rho_m}{dt} +3H(1+w_m)\rho_m & = & 0\\
\nonumber
\frac{d\rho_{rad}}{dt} +4H\rho_{rad} & = & 0
\end{eqnarray}
Since $w_0=-1$ then $\rho_0(t)=-p_0(t)$ at all $t$. The solutions of these entropic equations then are
\begin{eqnarray}
\rho_0(t) & = &\rho_{0,0}+\int \limits_{t_0}^t \frac{kT}{V}\frac{dS_0}{dt}=-p_0(t)\\
\nonumber
\rho_M(t) & = & \Bigl(\frac{a(t_0)}{a(t)}\Bigr)^3\Biggl[\rho_{M,0}+\int \limits_{t_0}^t \Bigl(\frac{a(t')}{a(t_0)}\Bigr)^{3}\frac{kT}{V}\frac{dS_M}{dt}\Biggr]\\
\nonumber
\rho_m(t) & = & \Bigl(\frac{a(t_0)}{a(t)}\Bigr)^{3(1+w_m)}\rho_{m,0}\\
\nonumber
\rho_{rad}(t) & = & \Bigl(\frac{a(t_0)}{a(t)}\Bigr)^4\rho_{rad,0}
\end{eqnarray}
The terms $\rho_m(t)$ and $\rho_M(t)$ play a dual role in our fits to the Hubble data~\cite{svec17b}. For $w_m=0$ there is no "residual" matter term and the terms $\rho_m(t)$ and $\rho_M(t)$ represent atomic matter and Dark Matter, respectively. For $w_m=w_r=-\frac{1}{3}$ the terms $\rho_m(t)$ and $\rho_M(t)$ represent the "residual" matter term and the combined Dark Matter + atomic matter term, respectively, in the extended Hubble function (3.30). 

These solutions must satisfy the cyclicity conditions $H(t_\alpha)=H(t_\omega)=0$ at the turning points $t_\alpha=0$ and $t_\omega=T/2$. It follows from (2.9) that these cyclicity conditions are satisfied when
\begin{eqnarray}
\bar{\rho}(t_\alpha) & = & \sum \limits_{k=0}^{rad} \rho_k(t_\alpha)=0\\
\nonumber
\bar{\rho}(t_\omega) & = & \sum \limits_{k=0}^{rad} \rho_k(t_\omega)=0
\end{eqnarray}
We shall refer to the equations (4.22) and (4.23) as the Special Entropic model. 

\subsection{Solutions for the entropies from the Gibbs-Duhem relations}

We now turn to solutions for the entropies $S_k$ from the Gibbs-Duhem relations (4.3). For Dark Energy $p_0=-\rho_0$ so that $\frac{dp_0}{dt}=-\frac{kT}{V}\frac{dS_0}{dt}$. We also have $\frac{dT}{dt}=-HT$ and 
$\frac{dS_0}{dt}=aH\frac{dS_0}{da}$, $\frac{d\mu_0}{dt}=aH\frac{d\mu_0}{da}$. Then the Gibbs-Duhem relation (4.3) for Dark Energy takes the form
\begin{equation}
\frac{dS_0}{da}-\frac{1}{a}S_0=-\frac{N_0}{kT}\frac{d\mu_0}{da}
\end{equation}
This equation can be integrated using (3.3) and then (3.9) to obtain
\begin{equation}
S_0(a)=\frac{a}{a_0}\Bigl[S_0(a_0)+\frac{a_0N_0}{kT_0}\frac{\mu_0(a_0)-\mu_0(a)}{a_0} \Bigr]
\end{equation}
For Dark Matter and atomic matter $p_M=0$ and $p_m=0$, respectively. For the radiation $p_{rad}=\frac{1}{3}\rho_{rad}$ and $\frac{dp_{rad}}{dt}=-\frac{4}{3}H\rho_{rad}$. Then the Gibbs-Duhem relations (4.3) for Dark Matter, atomic matter and radiation give
\begin{eqnarray}
S_M(a) & = & \Bigl(\frac{a}{a_0}\Bigr)^2\frac{a_0 N_M}{kT_0}\frac{d\mu_M}{da}\\
S_m(a) & = &  
\Bigl(\frac{a}{a_0}\Bigr)^2\frac{a_0 N_m(a)}{kT_0}\frac{d\mu_m}{da}\\
S_{rad}(a) & = & \Bigl(\frac{a}{a_0}\Bigr)^2\frac{a_0 N_{rad}(a)}{kT_0}\frac{d\mu_{rad}}{da}
+\frac{4}{3}\frac{\rho_{rad,0}V_0}{kT_0}
\end{eqnarray}
where $V_0$ is the present volume of the Universe. 

\subsection{Euler equations for the "residual" matter}

Galaxies are the basic building blocks of the Universe that evolve slowly. In the Section VII.D we use the kinetics of the space quanta forming such "static" galaxies to show that they possess an equation of state $P_i=-\frac{1}{3} \frac{\mathcal{E}_i}{V_i}$ where $\mathcal{E}_i$ and $V_i$ are the energy and volume of the galaxy $i$, respectively. Then we show that the entire ensemble of all such "static" galaxies is described by a similar equation of state
\begin{equation}
P_r=-\frac{1}{3} \frac{\mathcal{E}_r}{V_r}=-\frac{1}{3}\rho_r
\end{equation}
where $\mathcal{E}_r$, $V_r$ and $\rho_r$ are the total energy, volume and energy density of the entire ensemble of $N_r$ galaxies.  We refer to this equation of state as the "residual" matter term and identify the entire ensemble of the slowly evolving galaxies as the "residual" matter. This gives the "residual" matter a status of a new component of the Universe with the galaxies considered as its constituent "particles". 

We shall describe this fifth component of the Universe by Euler equations in the form
\begin{eqnarray}
\frac{d\rho_r}{dt}+3H\rho_r & = & -3HP_r+\frac{1}{V(t)} 
\biggl[kT(t)\frac{dS_r}{dt}+\mu_r(t)\frac{dN_r}{dt}\biggr]\\
V(t)\frac{dP_r}{dt} & = & \biggl[k\frac{dT}{dt}S_r(t) +
\frac{d\mu_r}{dt}N_r(t)\biggr]
\end{eqnarray}
subject to the constraint (4.31) and the requirement that
\begin{equation}
kT(t)\frac{dS_r}{dt}+\mu_r(t)\frac{dN_r}{dt}=0
\end{equation}
With $w_r=-\frac{1}{3}$ the solution of (4.32) then reads 
\begin{equation}
\rho_r(t)=\rho_{r,0}\Bigl(\frac{a_0}{a}\Bigr)^{3(1+w_r)}=\Bigl(\frac{a_0}{a}\Bigr)^3\rho_{r,0}\Bigl(\frac{a}{a_0}\Bigl)=\rho_{r,0}\Bigl(\frac{a_0}{a}\Bigr)^2
\end{equation}
Using $\frac{df(t)}{dt}=\frac{df(a)}{da}\frac{da}{dt}$, adding (4.33) and (4.34) and integrating we find
\begin{equation}
kT(a)S_r(a)+\mu_r(a)N_r(a)=
\frac{2}{3}\rho_{r,0}V(a_0)\Bigl(\frac{a}{a_0}\Bigr)
\end{equation}

The total entropy of the Universe is $S=S'+S''+S_r=const$ and its total particle number is $N=N'+N''+N_r=const$. We require that $S'=S_0+S_M=const$ and $S''=S_m+S_{rad}=const$ so that $S_r=const$. Similarly $N'=N_0+N_M=const$ and $N''=N_m+N_{rad}=const$ imply $N_r=const$. With the present volume of the Universe $V(a_0)=a^3_0\mathcal{V}$ where $\mathcal{V}$ is the comoving volume of the Universe, we can rewrite (4.36) in the form that defines the "residual" matter density $\rho_{r}(a)$
\begin{equation}
\frac{3}{2}\frac{kT_0a_0\sigma_r}{a_0^4}\biggl(\frac{a_0}{a}\biggr)^4 +\biggl(\frac{a_0}{a}\biggr)^3\frac{3}{2}\frac{n_r}{a_0^3}\mu_r(a)=\rho_{r,0}\biggl(\frac{a_0}{a}\biggr)^2=\rho_r(a)
\end{equation}
where the entropy and particle number densities 
$\sigma_r=\frac{S_r}{\mathcal{V}}$ and 
$n_r=\frac{N_r}{\mathcal{V}}$ are constants. 

We now show that with some redefinitions of the Dark Matter and radiation terms in the Hubble function (3.18) we recover the l.h.s. of the equation (4.37) and thus introduce the "residual" matter term directly into the Huble function. Carrying out the integration of (3.4) for $H^2$ with $\bar{p}$ given by (3.16) and using the relation (3.17) we find the "raw" form of the Hubble function
\begin{eqnarray}
H^2(a) & = & H_0^2\Bigg\{\biggl(\frac{8 \pi G}{3c^2H_0^2}\biggr)
\biggl[-\bar{p}_0 + \frac{kT_0a_0\sigma}{4a_0^4}+
\frac{n}{2}\frac{d\mu(\eta(a))}{da}\biggl(\frac{1}{a^2}-
\frac{1}{a^2_0}\biggr)\biggr]\\
 & + & \biggl(\frac{a_0}{a}\biggr)^3
 \biggl\{1+\biggl(\frac{8 \pi G}{3c^2H_0^2}\biggr) \biggl[\bar{p}_0-\frac{kT_0a_0\sigma}{a_0^4}+
 \frac{n}{a_0^3} \bigl(\mu(a)-\mu(a_0)\bigr)\biggr]\biggr\}\\
 & + & \biggl(\frac{8 \pi G}{3c^2H_0^2}\biggr)
 \frac{3}{4}\frac{kT_0a_0\sigma}{a_0^4}\biggl(\frac{a_0}{a}\biggr)^4 \Biggr\}
\end{eqnarray} 
Here $\sigma=\frac{S}{\mathcal{V}}=const$ and 
$n=\frac{N}{\mathcal{V}}=const$ are the total entropy and total particle number densities. We can write the two relevant terms in (4.39) and (4.40) in the form with an obvious notation 
\begin{eqnarray}
\frac{n}{a_0^3} \bigl(\mu(a)-\mu(a_0)\bigr) & = & 
\frac{n_M}{a_0^3} \bigl(\mu_M(a)-\mu_M(a_0)\bigr)+
\frac{3n_r}{2a_0^3} \bigl(\mu_r(a)-\mu_r(a_0)\bigr)\\
\frac{3}{4}\frac{kT_0a_0\sigma}{a_0^4}\biggl(\frac{a_0}{a}\biggr)^4 & = & 
\frac{3}{4}\frac{kT_0a_0\sigma_{rad}}{a_0^4}\biggl(\frac{a_0}{a}\biggr)^4 +
\frac{3}{2}\frac{kT_0a_0\sigma_r}{a^4}
\end{eqnarray}
where $\sigma=\sigma_{rad}+2\sigma_r$. In general there is no simple relation between $n$, $n_M$ and $n_r$. Assuming $n=n_M+\frac{3}{2}n_r$ implies the  identities $\mu_M(a)=\mu_r(a)\equiv \mu(a)$. Next we define $\Omega_{Mm,0}=\Omega_{M,0}+\Omega_{m,0}$ where
\begin{eqnarray}
\Omega_{Mm,0} & = & 1-\Omega_{0,0}-\Omega_{rad,0}-\Omega_{r,0}\\
\Omega_{0,0} & = & \biggl(\frac{8 \pi G}{3c^2H_0^2}\biggr) \biggl[-\bar{p}_0+\frac{kT_0a_0\sigma}{4a_0^4}\biggr]\\
\Omega_{rad,0} & = & \biggl(\frac{8 \pi G}{3c^2H_0^2}\biggr)
\frac{3}{4}\frac{kT_0a_0\sigma_{rad}}{a_0^4}\\
\Omega_{r,0} & = & \biggl(\frac{8 \pi G}{3c^2H_0^2}\biggr)
\biggl[\frac{3}{2}\frac{kT_0a_0\sigma_r}{a^4_0}
+\frac{3}{2}\frac{n_r}{a_0^3}\mu_r(a_0)\biggr]
\end{eqnarray}
Combining the $a$-dependent "residual" matter terms in (4.41) and (4.42) and using the result (4.37) with (4.35) we obtain the "residual" matter term in the  Hubble function (3.18)
\begin{equation}
\Omega_r(z)=\Omega_{r,0}(1+z)^{3(1+w_r)}
\end{equation}
It is noteworthy that in our best fits of this extended Hubble function to the Hubble data we take $\Omega_{r,0}=\Omega_{m,0}$ where $\Omega_{m,0}$ is the atomic matter term in fits with the "residual" matter term absent. Both fits yield identical $\chi^2/dof$~\cite{svec17b}.

\section{Quantum information theory of the Universe}

\subsection{Quantum Universe and its quantum spacetime}

Thermodynamics is a framework to describe how any system of very large numbers of individual contituent particles behave. We describe all four components of the Universe by Euler's equations. We thus implicitely assume that not only the atomic matter and radiation but also Dark Energy and Dark Matter have quantum structure consisting of distinct particles (quanta) of Dark Energy and Dark Matter, respectively.

The total entropies of the dark and visible sectors are both conserved during the evolution of the Universe. From the quantum point of view Dark Energy and Dark Matter ineract but form an isolated quantum system undergoing unitary evolution. Similarly the atomic matter and radiation interact but form an isolated quantum system undergoing unitary evolution. The internal quantum interactions within these two quantum systems are not gravitational interactios. These two quantum systems are coupled by gravitational interactions described by the Einstein's theory of gravity, specifically by the Friedmann equations. Supplemented by the Euler's equation they predict the existence of Dark Energy and Dark Matter in the Hubble function (3.18). 

Our direct observations suggest that the Universe consists of Space, atomic matter and radiation. This leads us to identify the dark sector of Dark Energy and Dark Matter with the Space itself. Then Space has a quantum structure and its constituents are the constituents of Dark Energy and Dark Matter. Standard Model of particle physics describes the quantum structure of the visible sector. In this Section we formulate an entropic model of the quantum structure of the dark sector. 

The homogeneous and isotropic background spacetime of the Cyclic Universe is a periodic gravitational wave $g_{\mu \nu}(\vec{x},t)$ given by the Robertson-Walker metric (2.1) with a periodic nonsingular scale factor $0<a_{min}\leq a(t)\leq a_{max}<\infty$. According to the principle of particle-wave duality this spacetime has a quantum structure. We model the spacetime quanta as two-qubit quantum states of two massless gravitons with two helicities $|\pm 2>$. A short review of two-qubit quantum states is given in the Appendix B and an important pedagogical review is given in  Ref.~\cite{friis16}.

All space quanta carry quantum information entropy~\cite{nielsen00,bengtsson06,vedral06}. The entangled quanta form Dark Energy while the non-entangled quanta form Dark Matter. While all quanta of Dark Energy carry quantum information entropy and entanglement entropy~\cite{vedral06}, all possible non-entangled quanta form Dark Matter. The entangled states of Dark Energy are non-local states that violate Bell inequality~\cite{bell64,bell87}.

Let $n(t,\rho_\lambda)$ be the probability distribution of the entangled states $\rho_\lambda$ at the time $t$. Then the expression
\begin{equation}
\rho_\lambda(t)=\int \limits_{\mathcal{M}_E} d\mu(\rho_\lambda)n(t,\rho_\lambda)\rho_\lambda
\end{equation}
defines the average quantum state of Dark Energy at the time $t$ assuming a measure $d\mu(\rho_\lambda)$ on the subspace of entangled states $\mathcal{M}_E$. 

The average entropy and the average entangement of space quanta forming the Dark Energy are  given by the entropy and entanglement of the quantum state $\rho_\lambda(t)$. This state is closely related to the scale factor $a(t)$ and the Robertson-Walker (RW) metric in cartesian coordinates (2.1). In the Fano basis the state $\rho_\lambda(t)=\sum \limits_{\mu,\nu=0}^3 t_{\mu \nu} \sigma^{\mu} \otimes \sigma^{\nu}$ is a density matrix given by
\begin{equation}
t_{00}=1, t_{11}=t_{22}=t_{33}=-\bar{a}(t)^2
\end{equation}
\[
t_{i 0}=t_{0j}=0, t_{ij}=0, i\neq j
\]
where $\bar{a}(t)=\frac{a(t)}{a_{max}}$ is the normalized scale factor. In the computational basis the nonzero components of 
$\rho_\lambda(t)=\sum \limits_{mn=00}^{11} \sum \limits_{m'n'=00}^{11}p_{mn,m'n'}|mn><m'n'|$ are given by
\begin{eqnarray}
p_{00,00}=p_{11,11} & = & \frac{1}{4}(1-\bar{a}(t)^2)\\
\nonumber
p_{01,01}=p_{10,10} & = & \frac{1}{4}(1+\bar{a}(t)^2)\\
\nonumber
p_{01,10}=p_{10,01} & = & -\frac{1}{2}\bar{a}(t)^2
\end{eqnarray}
The quantum information entropy $\Sigma_\lambda(t)$ carried by the state $\rho_\lambda(t)$ is given by von Neumann entropy $S(\rho)=-Tr(\rho \log \rho)$~\cite{nielsen00,bengtsson06,vedral06}
\begin{equation}
\Sigma_\lambda(t)=\frac{-1}{2\ln 2}\Bigl((1-\bar{a}^2)\ln(1-\bar{a}^2)
+(1+\bar{a}^2)\ln(1+\bar{a}^2)-2\bar{a}^2\ln(\frac{1}{2}\bar{a}^2)-
4\ln2 \Bigr)
\end{equation}
The entanglement content of the state $\rho_\lambda(t)$ is measured by the relative entropy of entanglement given by~\cite{vedral06}
\begin{equation}
\Sigma_E(t)=\min \limits_{\sigma \in \mathcal{M}_S} 
S(\rho_\lambda(t)||\sigma) = S(\rho_\lambda(t)||\sigma(t)) 
\end{equation}
where $S(\rho_\lambda(t)||\sigma(t))= +Tr(\rho_\lambda(t)\log\rho_\lambda(t))-Tr(\rho_\lambda(t)\log \sigma(t))=-\Sigma_\lambda(t)-Tr(\rho_\lambda(t)\log \sigma(t))>0$ is relative entropy and $\sigma(t)$ is the separable state nearest to the state $\rho_\lambda(t)$. It is defined by $Tr(\rho_\lambda(t)\log \sigma(t))= \min \limits_{\sigma \in \mathcal{M}_S}Tr(\rho_\lambda(t)\log \sigma)$. $\mathcal{M}_S$ is the subspace of separable states in the space of all two-qubit states $\mathcal{M}$. 

All possible separable (non-entangled) space quanta $\rho_\mu$ form Dark Matter with entropies $S(\rho_\mu)$. Let $p(t,\rho_\mu)$ be the probability distribution of the separable states $\rho_\mu$ at the time $t$. Then the expression
\begin{equation}
\rho_\mu(t)=\int \limits_{\mathcal{M}_S} d\mu(\rho_\mu)p(t,\rho_\mu)\rho_\mu
\end{equation}
defines the average quantum state of Dark Matter at the time $t$ assuming a measure $d\mu(\rho_\mu)$ on the subspace of separable states $\mathcal{M}_S$. This state carries a von Neumann entropy $\Sigma_\mu(t)$.

It is customary to calculate the von Neumann entropy $S(\rho)=-Tr(\rho \log \rho)$ for the diagonal form $\rho'=diag(\lambda_i)$ of the state $\rho$ where $\lambda_i,i=1,N$ are its non-zero eigenvalues. In this form the von Neumann entropy reads $S(\rho')=-\sum \limits_{i=1}^N \lambda_i \log \lambda_i$ and it is invariant under unitary transdormations. We have chosen to evaluate the entropy $\Sigma_\lambda(t)$ of Dark Energy using the general definition of the entropy $S(\rho)=-Tr(\rho \log \rho)$. 

Only atomic matter and radiation can interact to change their particles with a  change of particle numbers subject to $N_m(t)+ N_{rad}(t)=N''$. Since $N_0$ and $N_M$ are constant and the quanta of Dark Energy and Dark Matter are of two distinct kinds, Dark Energy and Dark Matter do not engage in mutual interactions or interactions with the atomic matter and radiation that would change their particle kind. However Dark Energy and Dark Matter do engage in entropy transfers that co-evolve their entropies. This interaction of Dark Energy and Dark Matter is described by the relation (3.29).

\subsection{Quantum duality relations for Dark Energy and Dark Matter on cosmological scales}

Dark Energy and Dark Matter are two distinct gravitational fields that cannot be detected by particle interactions of the Standard Model of particle physics. However, since these gravitational fields are two different forms of Space and the Space is a "material" component of the Universe, they should be both represented as two distinct terms in the Hubble function describing the evolution of the Universe. The Hubble function (3.18) of our General theory is such a Hubble function. 

The dual fields $h^E_{\mu \nu}$ and $h^M_{\mu \nu}$ of Dark Energy and Dark Matter both emerge from the quantum structure of the spacetime by quantum duality relations that define these fields. Baryonic matter (atomic matter and the radiation) is the source of gravitational field $h^B_{\mu \nu}$ and its relation to the quantum structure of spacetime is described by duality relations derived from the duality relations of Dark Matter and Dark Energy. 

In general, Dark Matter quantum states $\rho_\mu$ are described by a 
non-homogeneous probability distribution $p(\vec{x},t,S)$ of their quantum information entropy $S$. Gravity dual of Dark Matter quantum states $\rho_\mu$  is the gravitational field $g^M_{\mu\nu}$ defined by
\begin{equation}
g^M_{\mu\nu}=g^{(0)}_{\mu\nu}+h_{\mu\nu}^M
\end{equation}
The fields $h_{\mu\nu}^M$ are generated by Fisher information metric $F^M_{\mu\nu}$ of the probability distribution $p(\vec{x},t,S)$. The Fisher metric is defined by~\cite{amari93}
\begin{equation}
F^M_{\mu\nu}(\vec{x},t)=r^2_0\int \limits_0^2 dS p(\vec{x},t,S)
\frac{\partial \ln p(\vec{x},t,S)}{\partial x^\mu}
\frac{\partial \ln p(\vec{x},t,S)}{\partial x^\nu}
\end{equation}
where $r_0$ is a scale parameter and $x^0=ct$. Then the fields  $h_{\mu \nu}^M$ read
\begin{eqnarray}
h_{00}^M(\vec{x},t) & = & F_{00}(\vec{x},t)\\
\nonumber
h_{0j}^M(\vec{x},t) & = & a(t)F_{0j}(\vec{x},t)\\
\nonumber
h_{ij}^M(\vec{x},t) & = & a^2(t)F_{ij}(\vec{x},t)
\end{eqnarray}
The relations (5.8) and (5.9) constitute the quantum duality relations of Dark Matter. 

Gravity dual of Dark Energy quantum states $\rho_\lambda$ is the gravitational field $g^E_{\mu\nu}$ defined by
\begin{equation}
g^E_{\mu\nu}=g^{(0)}_{\mu\nu}+h_{\mu\nu}^E
\end{equation}
The fields $h^E_{\mu\nu}$ are generated by the Fisher information metric $F^E_{\mu\nu}$ of a probability distribution of variable(s) describing the quantum states $\rho_\lambda$. In general, Dark Energy quantum states $\rho_\lambda$ are described by a non-homogeneous probability distribution $n(\vec{x},t,S,S_E)$ of their quantum information entropy $S$ and entanglement entropy $S_E$ but the variables $S$ and $S_E$ are not independent variables.
They are related by the relation~\cite{vedral06} $S_E(\rho_\lambda)+S(\rho_\lambda)=\chi(\rho_\lambda)\geq 0$ where $\chi(\rho_\lambda)=Tr(\rho_\lambda\log \sigma_\rho)=\min \limits_{\sigma \in \mathcal{M}_S} Tr(\rho_\lambda \log \sigma)$ and $\sigma_\rho$ is the  separable state closest to the state $\rho_\lambda$. The distribution of Dark Energy $n(\vec{x},t,S,S_E)$ shows explicitely the dependence on both variables $S$ and $S_E$ but at the expense that the Fisher metric is a double integral with the integration over $S_E$ in the ill defined interval $[S_E(S)]$ of values of $S_E$ for a fixed value of $S$. However, for each $S_E\in[S_E(S)]$ there is a unique value of the variable $\chi=S_E+S\geq 0$. This means that the probabily distribution $n(\vec{x},t,S,S_E)$ is fully equivalent to a single variable distribution $q(\vec{x},t,\chi)$. This in turn means that we can write the Fisher metric of Dark Energy in the form
\begin{equation}
F^E_{\mu\nu}(\vec{x},t)=r^2_0\int \limits_0^{\chi_m} d\chi q(\vec{x},t,\chi)
\frac{\partial \ln q(\vec{x},t,\chi)}{\partial x^\mu}
\frac{\partial \ln q(\vec{x},t,\chi)}{\partial x^\nu}
\end{equation}
where $\chi_m=\text{Max}(\chi)$. The quantum duality relations for Dark Energy then read
\begin{eqnarray}
h_{00}^E(\vec{x},t) & = & F^E_{00}(\vec{x},t)\\
\nonumber
h_{0j}^E(\vec{x},t) & = & a(t)F^E_{0j}(\vec{x},t)\\
\nonumber
h_{ij}^E(\vec{x},t) & = & a^2(t)F^E_{ij}(\vec{x},t)
\end{eqnarray}
In general the fields $h^E_{\mu\nu}$ and $h^M_{\mu\nu}$ may not be small and the approximations $|h^E_{\mu\nu}|\ll 1$ and 
$|h^M_{\mu\nu}|\ll 1$ may not apply.

For the homogeneous and isotropic spacetime we specify the fields $g^E_{\mu\nu}(\vec{x},t)$ and $g^M_{\mu\nu}(\vec{x},t)$ in the Section VIII. Einstein's equations applied separately for the fields $g^E_{\mu \nu}$ and $g^M_{\mu \nu}$ define their energy-momentum stress tensors $T^E_{\mu \nu}(\vec{x},t)$ and $T^M_{\mu \nu}(\vec{x},t)$, respectively. Recall that the energy-momentum stress tensor $T^{RW}_{\mu \nu}(\vec{x},t)$ associated with the Robertson-Walker metric $g^{RW}_{\mu \nu}(\vec{x},t)$ has the form~\cite{weinberg08}
\begin{equation}
T^{RW}_{\mu \nu}(\vec{x},t):T^{RW}_{00}=\rho(t), T^{RW}_{i0}=0, T^{RW}_{ij}=p(t) g_{ij}(\vec{x},t)
\end{equation}
where $g_{ij}(\vec{x},t)$ are the spatial components of the Robertson-Walker metric given by (2.1). The energy-momentum stress tensors $T^E_{\mu \nu}$ and $T^M_{\mu \nu}$ have a similar form discussed in the Section IX.

\subsection{Dark Energy and Dark Matter in a local inertial frame}

\subsubsection{The prediction of the "residual" matter with $w_r=-\frac{1}{3}$}

In a local inertial frame with Minkowski metric Einstein equations in the linear approximation read~\cite{carroll04}
\begin{eqnarray}
G_{00} & = & 2 \nabla^2\Psi+\partial_k \partial_\ell s^{k\ell} =\frac{8\pi G}{c^2} T_{00}\\
\nonumber
G_{0j} & = & -\frac{1}{2}\nabla^2 w_j +\frac{1}{2}\partial_j \partial_k w^k +2\partial_0 \partial_j \Psi +\partial_0 \partial_k s_j^k=\frac{8\pi G}{c^2} T_{0j}\\
\nonumber
G_{ij} & = & (\delta_{ij}\nabla^2-\partial_i \partial_j)(\Phi-\Psi) 
+ \delta_{ij}\partial_0\partial_k w^k -\partial_0\partial_{(i}w_{j)} + 2\delta_{ij}\partial_0^2 \Psi\\
\nonumber
       &   & -(-\partial_0^2 +\nabla^2)s_{ij} + 2 \partial_k \partial_{(i}s_{j)}^k -\delta_{ij}\partial_k\partial_\ell s^{k\ell} =\frac{8\pi G}{c^2} T_{ij}
\end{eqnarray}
where $G_{\mu \nu}$ is the Einstein tensor and $s_{ij}=\frac{1}{2}(h_{ij}-\frac{1}{3}\delta^{kl}s_{kl}\delta_{ij})$. The trace over the spatial components is given by
\begin{equation}
TrG_{ij}=2\nabla^2(\Phi-\Psi)-\partial_0\partial_iw_i +6\partial_0^2 \Psi 
-\partial_k \partial_\ell s^{k\ell}= \frac{8\pi G}{c^2} Tr T_{ij}
\end{equation}
The general form of the momentum-energy stress tensor is given by~\cite{caprini18}
\begin{eqnarray}
T_{00} & = & \rho\\
\nonumber
T_{0j} & = & \partial_j u + u_j\\
\nonumber
T_{ij} & = & p\delta_{ij} +\bigl(\partial_i \partial_j -\frac{1}{3} \nabla^2\bigr)\sigma +\partial_iv_j +\partial_jv_i +\Sigma{ij}
\end{eqnarray}
The vector and tensor components satisfy the constraints
\begin{eqnarray}
\partial_i u_i & = & 0, \quad \partial_i v_i=0\\
\nonumber
\partial_i \Sigma_{ij} & = & 0, \quad \Sigma_{ii} = 0
\end{eqnarray}
From the conservation of the energy $\partial^\mu T_{\mu \nu}=0$ follow additional constraints
\begin{eqnarray}
\nabla^2 u & = & \dot{\rho}\\
\nonumber
\nabla^2 \sigma & = & \frac{3}{2}\bigl(\dot{u}-p\bigr)\\
\nonumber
\nabla^2 v_i & = & 2 \dot{u}_i
\end{eqnarray}
Notice that $T_{00}=\rho$ and $TrT_{ij}=3p$.

In the linear approximation the fields $|h^M_{\mu \nu}|\ll 1$ and 
$|h^E_{\mu \nu}| \ll 1$ and can be expressed in terms of the gravitational  potentials.  Then the perturbations $h_{\mu \nu}^M$ read~\cite{carroll04,lyth09}
\begin{eqnarray}
h_{00}^M(\vec{x},t) & = & F_{00}(\vec{x},t)=-2\Phi^M\\
\nonumber
h_{0j}^M(\vec{x},t) & = & a(t)F_{0j}(\vec{x},t)=aw_j^M\\
\nonumber
h_{ij}^M(\vec{x},t) & = & a^2(t)F_{ij}(\vec{x},t)=a^2\bigl(-2\Psi^M \delta_{ij}+2s_{ij}^M\bigr)
\end{eqnarray}
where $\Psi^M=-\frac{1}{6}Tr h_{ij}^M$ and $s_{ij}^M$ is traceless. 
Relations for Dark Energy are similar.

Consider a local region $R$ in this frame of a free space devoid of Baryonic matter, such as a cosmic void. The local Dark Matter $h^{MV}_{\mu \nu}$ and Dark Energy $h^{EV}_{\mu \nu}$ are given by (5.9) and (5.12),respectively, with $a(t)=1$. Using the Einstein equations (5.14) these two fields define local energy-momentum stress tensors $T^{MV}_{\mu \nu}$ and $T^{EV}_{\mu \nu}$. 

Next, consider the same region of space filled with Baryonic matter such as a galaxy, a cluster or even a supercluster of galaxies. The presence of the Baryonic matter will displace the probability distributions $p \rightarrow p'$ and $q \rightarrow q'$ leading to new fields $h^{MB}_{\mu \nu}$ and $h^{EB}_{\mu \nu}$ of Dark Matter and Dark Energy which define new energy-momentum tensors $T^{MB}_{\mu \nu}$ and $T^{EB}_{\mu \nu}$. 

The spatial traces of $T^{MV}_{\mu \nu}$ and $T^{EV}_{\mu \nu}$ are equal to the pressure terms $3p^{MV}$ and $3p^{EV}$ while the $00$ components of these tensors define energy the densities $\rho^{MV}$ and $\rho^{EV}$. Similarly, the spatial traces of $T^{MB}_{\mu \nu}$ and $T^{EB}_{\mu \nu}$ are given by the pressure terms $3p^{MB}$ and $3p^{EB}$ while the $00$ components of these tensors define the energy densities $\rho^{MB}$ and $\rho^{EB}$. 

Consider now static distributions $p$ and $p'$ of Dark Matter on such galactic scales. Then the $00$ and $0j$ components of the fields $h^{MV}_{\mu \nu}$ and $h^{MB}_{\mu \nu}$ are all equal to zero. The Einstein equations (5.15) then imply in both these cases that $Tr G_{ij}=-G_{00}$ so that $Tr T_{ij}=-T_{00}$. Hence we have
\begin{eqnarray}
p^{MV} & = & -\frac{1}{3}\rho^{MV}\\
\nonumber
p^{MB} & = & -\frac{1}{3}\rho^{MB}
\end{eqnarray}
The equations of state for both these cases are equal to $w_r=-\frac{1}{3}$. Similarly we find for the Dark Energy on the same galactic scales
\begin{eqnarray}
p^{EV} & = & -\frac{1}{3}\rho^{EV}\\
\nonumber
p^{EB} & = & -\frac{1}{3}\rho^{EB}
\end{eqnarray}
For $r \to \infty$ we expect vanishing volume averages for radially symmetric galaxies
\begin{equation}
<\rho^{MB}>=<p^{MB}>=0
\end{equation}
Similarly the other averages to vanish in this limit. Thus there are no cosmological contributions with equations of state $-\frac{1}{3}$ from such Dark Matter and Dark Energy. However, the real  galaxies are finite in their extent and may not be radially symmetric. Then for $k=MB,EB$ the volume averages do not exactly vanish and their residuals are given by
\begin{equation}
<p^k>=-\frac{1}{3}<\rho^k>
\end{equation}
On cosmological scales these contributions from Dark Matter and Dark Energy on galactic systems combine (see equations (7.29) and (7.30)) to produce a cosmological "residual" matter with the density
\begin{equation}
\rho_r(t)=\rho_{r,0}\Bigl(\frac{a(t_0)}{a(t)}\Bigr)^{3(1+w_r)}=\rho_{r,0}(1+z)^{3(1+w_r)}=\rho_{r,0}(1+z)^2
\end{equation}
with the equation of state $w_r=-\frac{1}{3}$. This term originates entirely from the presence of the Dark Matter and Dark Energy. This result is a unique prediction of our model of the quantum structure of spacetime that is testable in fits to the Hubble data $H(z)$. The fits confirm its existence and suggest that $\rho_{r,0}=\rho_{m,0}$~\cite{svec17b}.

\subsubsection{The prediction of the negative spatial curvature}

Because the definition of the Hubble parameter involves only the scale factor and not the curvature parameter $R_0$ in the RW metric, the curvature term is absent in the Friedmann equations (2.9) and (2.10). We can still define fractional curvature density
$\Omega_{c}(t)=\rho_c(t)/\bar{\rho}(t)$ so that
\begin{equation}
\Omega_{c,0}=\frac{-kc^2}{R_0^2 H_0^2 a_0^2}
\end{equation}
where $a_0=a(t_0)$. The new term $\rho_r(z)=\rho_{r,0}(1+z)^2$ in (5.24) is akin to the curvature term $\rho_c(z)=\rho_{c,0}(1+z)^2$  with $k=-1$. Since the Hubble function does not depend on the  curvature term $\rho_c(z)$, the new term $\rho_r(z)$ is distinct from $\rho_c(z)$. Since $\rho_r(z)$ mimics $\rho_c(z)$ it can be interpreted as a curvature of the space internally generated by Dark Matter and Dark Energy additional to the spatial curvature $\rho_c(z)$ of the RW metric. It is this internal curvature $\rho_r(z)$ that contributes to the Hubble function and thus participates in the evolution of the Universe while the external curvature $\rho_c(z)$ does not participate. 

The consistency of the internal curvature $\rho_r(z)$ and the external curvatures $\rho_c(z)$ requires that they have the same sign. The positivity of $\rho_r(z)$ then suggests that Hubble data select $k=-1$. This means the spacetime of our Universe is anti-de Sitter spacetime with the curvature density $\Omega_{c,0}>0$. This is another  prediction of our model of the quantum structure of spacetime which is testable in fits of the Hubble function to luminosity distance or angular diameter distance data.

\subsubsection{The implications of negative spatial curvature for cosmological parameters}

The predicted positivity of the curvature density $\Omega_{c,0}$ has significant observable consequences for the cosmological parameters $\Omega_{0,0}$ and $\Omega_{Mm,0}$ of Dark Energy and Dark Matter. The luminosity distance $d_L(z)$ is given by the relation which explicitely depends on $\Omega_{c,0}$\cite{weinberg08,carroll04} 
\begin{equation}
d_L(z) = (1+z)\frac{1}{\sqrt{|\Omega_{c,0}}|}S_k
\Bigl[\sqrt{|\Omega_{c,0}}|\int \limits_0^z \frac{dz'}{H(z')} \Bigr]\\
\end{equation}
where $S_k(\chi)= \sin(\chi), \chi,\sinh(\chi)$ apply to closed, flat and open geometry with $k=+1,0,-1$, respectively. The luminosity distance relation is a direct consequence of the homogeneity and isotropy of the spacetime~\cite{carroll04}. Taking the first and second derivatives of $d_L(z)$ and combining the results we find a linear differential equation for $H^2$ with a solution for $\Omega_{c,0}$
\begin{equation}
\Omega_{c,0}=\frac{H^2 y^{'2}-c^2}{H_0^2y^2}=const
\end{equation}
where $y(z)=\frac{d_L(z)}{1+z}$ and $c$ is the speed of light. This relation was first derived in Ref.~\cite{clarkson07}. It is evident from (5.26) and (5.27) that $\Omega_{c,0}$ depends crucially on the choice of the Hubble function $H(z)$ for the fixed cosmological data on $y(z)$.

It is well known that the fits of the $\Lambda$CDM Model to the luminosity distance data are consistent with $\Omega_{c,0}=0$~\cite{planck15}. The Hubble function of the $\Lambda$CDM Model is given by
\begin{equation}
H^2_{\Lambda CDM}=H_0^2\Bigl[\Omega_\Lambda +(1+z)^3 \Omega_{Mm,0}^\Lambda +(1+z)^4\Omega_{rad,0}^\Lambda \Bigr]
\end{equation}
With $H^2=H^2_{\Lambda CDM}$ in (5.27) we find 
$y^{'2}=\frac{c^2}{H^2_{\Lambda CDM}}$ so that (5.27) reads
\begin{equation}
\Omega_{c,0}=\frac{H^2-H^2_{\Lambda CDM}}{H_0^2}\Bigl(\frac{y'}{y} \Bigr)^2
\end{equation}
The Hubble function $H^2(z)$ in the Entropic Model has a general form
\begin{equation}
H^2=H_0^2\Bigl[\Omega_{0,0}+(1+z)^3\bigl(\Omega_{Mm,0}+\Omega_{r,0}\bigr)+(1+z)^4\Omega_{rad,0}+\Sigma_0+(1+z)^3 \Sigma_M \Bigr]
\end{equation}
At $z=0$ both Hubble functions coincide $H^2=H^2_{\Lambda CDM}=H_0^2$. Neglecting the same or very similar small terms $\Omega_{rad,0}$ and $\Omega_{rad,0}^\Lambda $~\cite{PDG2015,PDG2014} we have a condition
\begin{equation}
\Omega_{0,0}+\Omega_{Mm,0}+\Omega_{r,0}=
\Omega_\Lambda + \Omega_{Mm,0}^\Lambda 
\end{equation}
Then at small $z$ the condition $\Omega_{c,0}>0$ implies
\begin{equation}
H^2-H^2_{\Lambda CDM}=3z\bigl(\Omega_{Mm,0}+\Omega_{r,0}-
\Omega_{Mm,0}^\Lambda \bigr) +\Sigma_0+(1+3z)\Sigma_M >0
\end{equation}
Integrating the condition (3.29) at small $z$ using (3.13) we obtain
\begin{equation}
\int \Bigl(\frac{d\Sigma_0}{dz}+(1+3z)\frac{d\Sigma_M}{dz}\Bigr)dz=
\Sigma_0(z)+(1+3z')\Sigma_M(z)=0
\end{equation}
where $0<z'<z$. Then for small $z \approx z'$ also $\Sigma_0(z)+(1+3z)\Sigma_M(z)\approx 0$ and the conditions (5.32) and (5.31) imply
\begin{eqnarray}
\Omega_{Mm,0}+\Omega_{r,0} & > & \Omega_{Mm,0}^\Lambda\\
\Omega_{0,0} & < & \Omega_\Lambda
\end{eqnarray}
Because of the presence of the entropic terms $\Sigma_0$ and $\Sigma_M$ in the entropic Hubble function we expect the fits to the Hubble data yield $\Omega_{Mm,0}+\Omega_{r,0} \neq \Omega_{Mm,0}^\Lambda$ and $\Omega_{0,0} \neq \Omega_\Lambda$. The inequalities (5.34) and (5.35) are specific consequences of $\Omega_{c,0}>0$ testable in fits of $H^2$ to the Hubble data. They are thus a unique signature of our model of Dark Energy and Dark Matter as gravitational fields arising from the quantum structure of spacetime in terms of the Fisher metric duality relations.

\subsection{Quantum duality relations for the Baryonic matter}

The Baryonic gravitational field $h^B_{\mu\nu}(\vec{x},t)$ is sourced by the local energy momentum tensor $T^B_{\mu \nu}(\vec{x},t)$. The dual gravitational fields of Dark Energy and Dark Matter constitute the Space. We can consider the Baryonic gravitational field as its perturbation. The gravitational fields of Dark Matter and Dark Energy are given by the quantum duality relations relating the two fields to the quantum structure of the Space. We now seek a duality relation for the Baryonic gravitational field relating it to its perturbation of the quantum structure of the Space.

In Classical Electrodynamics the macroscopic electromagnetic fields in a free space with no matter are the electric field $\vec{E}$ and the magnetic field $\vec{B}$. In the presence of a material medium the displaced electric field $\vec{D}$ and magnetic field $\vec{H}$ are given by~\cite{jackson75}
\begin{eqnarray}
\vec{D}=\vec{E}+4 \pi \vec{P}, & \quad & \vec{P}=\vec{P}(\vec{E},\vec{B})\\
\vec{H}=\vec{B}+4 \pi \vec{M}, & \quad & \vec{M}=\vec{M}(\vec{E},\vec{B})
\end{eqnarray}
where $\vec{P}$ and $\vec{M}$ are the polarization and magnetization of the material medium, respectively. In a linear and isotropic medium 
\begin{equation}
\vec{P}=\chi_e \vec{E}, \quad \vec{M}=\chi_m\vec{B}
\end{equation}
where the scalars $\chi_e$ and $\chi_m$ are electric and magnetic susceptibilities of the medium, respectively.

Taking an inspiration from the Electrodynamics we assume
\begin{eqnarray}
h^{MB}_{\mu \nu}=h^{MV}_{\mu \nu}+ 4 \pi P_{\mu \nu}, & \quad & 
P_{\mu \nu}=\chi_M h^{MV}_{\mu \nu}\\
h^{EB}_{\mu \nu}=h^{EV}_{\mu \nu}+ 4 \pi M_{\mu \nu}, & \quad & 
M_{\mu \nu}=\chi_E h^{EV}_{\mu \nu}
\end{eqnarray}
where we shall assume 
\begin{equation}
\chi_{M}=\chi_M^0 h^B_{\alpha \beta}\eta^{\alpha \beta}, \quad 
\chi_E=\chi_E^0 h^B_{\alpha \beta}h^{B,\alpha \beta}, 
\quad |\chi_E| \ll |\chi_M|
\end{equation}
With $p'=p+\delta p$ we find from $h^{MB}_{\mu \nu}-h^{MV}_{\mu \nu}$ the First quantum duality relation for Baryonic fields 
\begin{eqnarray}
4 \pi P_{\mu \nu} & = & 4 \pi \chi_M^0 h^B_{\alpha \beta} \delta^{\alpha \beta} h^{MV}_{\mu \nu}\\
\nonumber
  & = & -r_0^2 \int \limits_0^2 dS \frac{1}{p}\frac{\delta p}{p}
  \frac{\partial p}{\partial x^\mu}\frac{\partial p}{\partial x^\nu}
+r_0^2 \int \limits_0^2 dS\Bigl(1-\frac{\delta p}{p}\Bigr) \frac{1}{p}\Biggl\{
\frac{\partial p}{\partial x^\mu}\frac{\partial \delta p}{\partial x^\nu}+
\frac{\partial \delta p}{\partial x^\mu}\frac{\partial p}{\partial x^\nu}+
\frac{\partial \delta p}{\partial x^\mu}\frac{\partial \delta p}{\partial x^\nu}\Biggr\}
\end{eqnarray}
With $q'=q+\delta q$ we find from $h^{EB}_{\mu \nu}-h^{EV}_{\mu \nu}$ the Second quantum duality relation for Baryonic fields
\begin{eqnarray}
4 \pi M_{\mu \nu} & = & 4 \pi \chi_E^0 h^B_{\alpha \beta}h^{B,\alpha \beta} h^{EV}_{\mu \nu}\\
\nonumber
  & = & -r_0^2 \int \limits_0^{\chi_m} d\chi 
  \frac{1}{q}\frac{\delta q}{q}
  \frac{\partial q}{\partial x^\mu}\frac{\partial q}{\partial x^\nu}\\
\nonumber  
  &  & 
+r_0^2 \int \limits_0^{\chi_m} d\chi 
\Bigl(1-\frac{\delta q}{q}\Bigr) \frac{1}{q}\Biggl\{
\frac{\partial q}{\partial x^\mu}\frac{\partial \delta q}{\partial x^\nu}+
\frac{\partial \delta q}{\partial x^\mu}\frac{\partial q}{\partial x^\nu}+
\frac{\partial \delta q}{\partial x^\mu}\frac{\partial \delta q}{\partial x^\nu}\Biggr\}
\end{eqnarray}
We can express the fields $h^{MV}_{\mu \nu}$ and $h^{EV}_{\mu \nu}$ in terms of the Fisher metric integrals and bring the terms on the l.h.s. of (5.42) and (5.43) to the r.h.s. under the same integrations to obtain integro-differential equations for $\delta p$ and $\delta q$ of the form $0=\int F(\vec{x},t,S)dS$ and $0=\int G(\vec{x},t,\chi)d\chi$, respectively. Sufficient conditions for these equations to hold at all $\vec{x},t$ are the differential equations $F(\vec{x},t,S)=0$ and $G(\vec{x},t,\chi)=0$ for all values of $S$ and $\chi$, respectively. Assuming that the probability densities $p(\vec{x},t,S)$ and $q(\vec{x},t,\chi)$ of the free space are known, these equations can be solvable in a linear approximation of $\delta p(\vec{x},t,S)$ and $\delta q(\vec{x},t,\chi)$ to determine these perturbations from the given Baryonic gravitational field $h^B_{\mu \nu}$. These relations are the Baryonic quantum duality relations. In principle the distribution functions $p(\vec{x},t,S)$ and $q(\vec{x},t,\chi)$ can be inferred from the study of cosmic voids or from some theoretical principles.

\subsection{Simplified forms of the probabilty distributions of Dark Matter and Dark Energy}

The task of solving the duality relation (5.42) for $\delta p$ in terms of the Baryonic field can be greatly simplified. The finite range $0\leq S \leq 2$ of the entropy of Dark Matter quanta suggests the probability density $p(\vec{x},t,S)$ is given by the normalized Beta function distribution~\cite{tijms04} 
\begin{equation}
\tilde{p}(x,\alpha,\beta)= \frac{1}{B(\alpha,\beta)}x^{\alpha-1}(1-x)^{\beta -1}
\end{equation}
where $0\leq x \leq 1$, $\alpha>0, \beta >0$ and the Beta function
\begin{equation}
B(\alpha,\beta)=\frac{\Gamma(\alpha)\Gamma(\beta)}{\Gamma(\alpha+\beta)}
\end{equation}
The expected value and the variance of the random variable $X$ are given by~\cite{tijms04}
\begin{eqnarray}
E(X) & = & \frac{\alpha}{\alpha+\beta}\\
\nonumber
Var(X) & = & \frac{\alpha \beta}{(\alpha+\beta)^2 (\alpha+\beta +1)}
\end{eqnarray}
We set $p(\vec{x},t,S)=p(\vec{x},t,2x)=\tilde{p}(x,\alpha,\beta)$ and $\beta=1$. With $\Gamma(\alpha+1)=\alpha \Gamma(\alpha)$ and $\Gamma(1)=1$ we define the normalized probability density 
\begin{equation}
\tilde{p}(\vec{x},t,x)=\alpha(\vec{x},t)x^{\alpha(\vec{x},t)-1}
\end{equation}
where $\alpha(\vec{x},t)\geq1$ describes the dependence of $p(\vec{x},t,S)$ on the position and time. For $\alpha(\vec{x},t)=1$ the probabilty $\tilde{p}(\vec{x},t,x)=1$ for all entropy values, including pure states with $x=S=0$. Calculating the derivatives and carrying out the intergrations (see Ref.~\cite{svec17c}) we obtain an elegant result for the duality relation (5.9) (with $a(t)=1$)
\begin{equation}
h^{MV}_{\mu \nu} = r^2_0\frac{2}{\alpha^2}\frac{\partial \alpha}{\partial x^\mu}\frac{\partial \alpha}{\partial x^\nu}
\end{equation}
The perturbation $\delta \bar{p}= \delta \Bigl(\alpha x^{\alpha-1}\Bigr)=\frac{\delta \alpha}{\alpha}
(1+\ln x^\alpha)\bar{p}$ defines the Baryonic duality relation. In this picture the functions $\alpha(\vec{x},t)$ and $\delta
\alpha(\vec{x},t)$ are effectively the quantum structure duals of the gravitational fields $h^{MV}_{\mu \nu}(\vec{x},t)$ and $h^B_{\mu \nu}(\vec{x},t)$, respectively. The expected value of the entropy of the Dark Matter and its variance are given by
\begin{eqnarray}
\bar{S}_M(\vec{x},t) & = & 4\frac{\alpha(\vec{x},t)}{\alpha(\vec{x},t)+1}\\
\nonumber
(\delta S_M(\vec{x},t))^2 & = & 8\frac{\alpha(\vec{x},t)}{(\alpha(\vec{x},t)+1)^2(\alpha(\vec{x},t)+2)}
\end{eqnarray}

Next we shall consider a static and radially symmetric probability density $\bar{p}(r,x)$. Then $h^{MV}_{00}=h^{MV}_{0j}=0$ and $P_{00}=P_{0j}=0$. With the notation $A=\frac{r_0}{\alpha}\frac{d \alpha}{dr}$ the duality relations (5.48) and (5.42) read 
\begin{eqnarray}
h^{MV}_{ij} & = & 2A^2 \Omega_i \Omega_j\\
4\pi P_{ij} & = & 4\pi \chi_M 2A^2 \Omega_i \Omega_j\\
\nonumber
  & = & \Biggl\{
  2\Bigl(1+2\frac{\delta \alpha}{\alpha}\Bigr)
  \Bigl(\frac{r_0}{\alpha}\frac{d\delta \alpha}{dr}\Bigr)^2
  +4A\Bigl(\frac{r_0}{\alpha}\frac{d\delta \alpha}{dr}\Bigr)
  - 8A^2\Bigl(\frac{\delta \alpha}{\alpha}\Bigr)
  -16A^2\Bigl(\frac{\delta \alpha}{\alpha}\Bigr)^2
  +80A^2\Bigl(\frac{\delta \alpha}{\alpha}\Bigr)^3 
  \Biggr\}\Omega_i \Omega_j
\end{eqnarray}  
Here $\Omega_1=\cos \phi \sin \theta$, $\Omega_2=-\sin \phi \cos \theta$, $\Omega_3=\cos \theta$ where $\phi, \theta$ are polar coordinates. In the linear approximation of (5.51) the solution for $\delta \alpha$ reads
\begin{equation}
\delta \alpha = \Bigl(\frac{\alpha}{\alpha_i}\Bigr)^2\Biggl\{\delta \alpha_i+2\pi \int \limits_{r_i}^r \Bigl(\frac{\alpha_i}{\alpha}\Bigr)^2
\frac{d\alpha}{dr} \chi_M(r) dr \Biggr\}
\end{equation}
For static radially symmetric Baryonic field $h^B_{\mu \nu}=\text{diag}(-2\Phi,-2\Phi,-2\Phi,-2\Phi)$ where $\Phi(r)=-\frac{GM}{r}$ is the scalar potential~\cite{carroll04}. Then $\chi_M=-4\chi_M^0\Phi$ and we have a solution for the perturbation $\delta p(r,S)$ in terms of the Baryonic field. Notice that $|\chi_E|=16|\chi_E^0|\Phi^2\ll|\chi_M|$.

We can write the Fisher metric of Dark Energy in a general form
\begin{equation}
F^E_{\mu\nu}(\vec{x},t)=r^2_0\int \limits_0^{\chi_m} d\chi q(\vec{x},t,\chi)
\frac{\partial \ln q(\vec{x},t,\chi)}{\partial x^\mu}
\frac{\partial \ln q(\vec{x},t,\chi)}{\partial x^\nu}
\end{equation}
where $\chi_m=\text{Max}(\chi)$. To render Dark Energy similarly tractable we can define $q(\vec{x},t,\chi)$ as a Beta function distribution
\begin{equation}
q(\vec{x},t,\chi)=\tilde{q}(\vec{x},t,x)=
\beta(\vec{x},t)x^{\beta(\vec{x},t)-1}
\end{equation}
where $x=\frac{\chi}{\chi_m}$ and $\beta>1$. For pure states with $x=\chi=0$ the probabilty $\tilde{q}(\vec{x},t,0)=0$. To distinguish the Beta fuction of Dark Energy from that of Dark Matter we use the symbol $\beta(\vec{x},t)$ to represent the variable $\alpha(\vec{x},t)$ in the definition of the Beta function.  

Using Fisher metric (5.53) the gravitational field of Dark Energy is given by (with $a(t)=1)$
\begin{equation}
h^{EV}_{\mu \nu} = r^2_0\frac{\chi_m}{\beta^2}\frac{\partial \beta}{\partial x^\mu}\frac{\partial \beta}{\partial x^\nu}
\end{equation}
The perturbation $\delta \bar{q}= 
 \delta \Bigl(\beta x^{\beta-1}\Bigr)=
 \frac{\delta \beta}{\beta}(1+\ln x^\beta)\bar{q}$ 
defines the Baryonic duality relation. The solution for $\delta \beta$ for static and radially symmetric $q(r,\chi)$ in linear approximation is given by an equation similar to (5.52). The expected value and the variance of the variable $\chi$ of the Dark Energy are given by expressions similar to (5.49)
\begin{eqnarray}
\bar{\chi}(\vec{x},t) & = & \chi_m^2\frac{\beta(\vec{x},t)}{\beta(\vec{x},t)+1}\\
\nonumber
(\delta \chi(\vec{x},t))^2 & = & \chi_m^3\frac{\beta(\vec{x},t)}{(\beta(\vec{x},t)+1)^2(\beta(\vec{x},t)+2)}
\end{eqnarray}

In conclusion, we propose two dual descriptions of the Space:\\
(1) a microscopic description in terms of the two probability densities $p, \delta p$ and $q$, $\delta q$\\
(2) a macroscopic description in terms of the two gravitational fields $h^{EV}_{\mu \nu},h^{MV}_{\mu \nu}$ and $h^{EB}_{\mu \nu},h^{MB}_{\mu \nu}$.

\section{The total entropy of Dark Energy and Dark Matter.}

\subsection{Transformations of the total entropies in the Special Entropic Theory}

The total entropy $S_0(t)$ of Dark Energy is a functional $S_0(\Sigma_\lambda(t))$ of the entropy $\Sigma_\lambda(t)$. At the turning points we have 
\begin{eqnarray}
\bar{a}(t=0)=\bar{a}_{min}>0 & => & \Sigma_\lambda(\bar{a}_{min})=\Sigma_{\lambda,max}\sim 2\\
\nonumber
\bar{a}(t=T/2)=\bar{a}_{max}=1 & => & \Sigma_\lambda(\bar{a}_{max})=\Sigma_{\lambda,min}=0\\
\nonumber
\bar{a}(t=T)=\bar{a}_{min}>0 & => & \Sigma_\lambda(\bar{a}_{min})=\Sigma_{\lambda,max}\sim 2
\end{eqnarray}
During the expansion phase the entropy of Dark Energy decreases from $S_0(\Sigma_{\lambda,max})=S_{0,max}=S'$ at $t=0$ to $S_0(\Sigma_{\lambda,min})=S_{0,min}=0 $ at $t=T/2$. During the contraction phase it increases back to $S_{0,max}=S'$ at $t=T$. 

Recall that $S_0(t)+S_M(t)=S'=const$. This relation implies that we can view the total entropy of Dark Matter also as a functional $S_M(\Sigma_\lambda(t))$. During the expansion entire entropy $S_{0,max}$ at $t=0$ is transfered to Dark Matter the entropy of which increases from $S_{M,min}=0$ at $t=0$ to $S_{M,max}=S'$ at $t=T/2$. During the contraction Dark Matter gives it all back to Dark Energy at $t=T$ to start a new cycle. The cycles are governed by the dependence of $\Sigma_\lambda(t)$ on the periodic scale factor $a(t)$.

In addition to its quantum information entropy $S_0(t)$ Dark Energy 
carries entanglement entropy $S_E(t)=S_E(\Sigma_E(t))$ arising from the  entanglement entropy $\Sigma_E(t)$ of its quantum states $\rho_\lambda(t)$. We relate the entropy $S_E(t)$ to the cosmological pressure $p_0(t)$ of Dark Energy
\begin{equation}
-3Hp_0=\frac{kT}{V}\frac{dS_E}{dt}
\end{equation}
Since Dark Matter carries no entanglement its cosmological pressure $p_M=0$.

It follows from (6.2) that $\frac{dS_E}{dt}=0$ at the turning points. Moreover, the pressure $p_0=0$ at $t=t_p$ where it changes sign. If during the expansion $p_0$ is negative for $t<t_p$ and positive for $t>t_p$ then the entanglement entropy is increasing from some minimum $S_{E,min}>0$ at $t=t_{1}<t_p$ to a maximum $S_{E,max}$ at $t=t_p$ and then decreases back to some minimum $S_{E,min}>0$ at $t=t_2>t_p$. The vanishing of $S_E$ at the turning points is exluded because Dark Energy quanta must be entangled. In contrast, the entropy $S_0$ can vanish at $t=T/2$ indicating entangled pure states. Similarly the entropy $S_M$ can vanish at $t=0$ or $t=T$ indicating pure non-entangled states. 

Because the total entropy $S=S'+S''$ of the Universe is constant, the total entropy of the atomic matter and radiation is constant $S_m(t)+S_{rad}(t)=S''$. We can assume that during the expansion the total entropy of radiation $S_{rad}(t)=S_{rad}(\rho_{rad}(t))$ decreases from its maximum value $S_{rad,max}=S''$ at $t=0$ to its minimum value $S_{rad,min}=0$ at $t=T/2$. At the same time the entropy of the atomic matter $S_m(t)=S''-S_{rad}(t)$ increases from a minimum $S_{m,min}=0$ at $t=0$ to a maximum $S_{m,max}=S''$ at $t=T/2$. During the contraction the process is reversed with the entropy of the atomic matter fully returned to the radiation. This cyclic process mirrors the similar cyclic process in the dark sector. There is no entropic death of the Universe. 

Finally we note that in addition to the entropic equations (4.20) and (4.21) for atomic matter and radiation, respectively, the relation (4.19) implies a dynamical relation
\begin{equation} 
\frac{\mu_m}{V}\frac{dN_m}{dt}+\frac{\mu_{rad}}{V}\frac{dN_{rad}}{dt}=0
\end{equation}
Combining this relation with $N''=N_m(t)+N_{rad}(t)$ we find a condition  $(\mu_m-\mu_{rad})\frac{dN_m}{dt}=0$ with two solutions: (i) $\mu_m=\mu_{rad}$ and (ii) $\frac{dN_m}{dt}=\frac{dN_{rad}}{dt}=0$. The first solution implies transformations of the particles of radiation into particles of matter and vice versa during the evolution with $\frac{dN_m}{dt}=-\frac{dN_{rad}}{dt}$ associated with the transformations of the entropy $\frac{dS_m}{dt}=-\frac{dS_{rad}}{dt}$. There are no such transformations allowed in the second solution with $\frac{dN_m}{dt}=\frac{dN_{rad}}{dt}=0$ and $\frac{dS_m}{dt}=\frac{dS_{rad}}{dt}=0$. The second solution is disallowed because the transformations of radiation into atomic matter and vice versa during the evolution do arise from their particle interactions and particle redshifts (e.g. neutrinos). 
  
We recognize the terms $kTdS_k= dQ_k$ defining the energy densities in (4.7) as the energy equivalent of the information entropy first proposed by Landauer in 1961 for the erasure of information in computation processes~\cite{landauer61}. Conversion of quantum information entropy of a single quantun system into energy was recently experimentally confirmed~\cite{peterson16,yan18}. In our picture the quantum information of Dark Energy (radiation) is erased during the expansion and stored in the Dark Matter (atomic matter). During the contraction the process is reversed. The entropy of atomic matter $S_m(t)$ includes the total entropy of the increasing complexity of its self-organizing structures~\cite{prigogine77,jantsch80,kauffman93} emerging during the expansion, including chemical evolution and the evolution of Life and consciousness, and reaching an apex at $t_\omega$. On a fundamental level the complexity of a structure and its function is related to the level of its information content and its processing power.

\subsection{Quantum model of the total entropies $S_0(t)$ and $S_M(t)$}

With the final solutions for the energy densities (4.22) the Hubble function in the Entropic theory reads
\begin{equation}
H^2=H_0^2\Bigl[\Omega_{0,0}+\Sigma_0(z)+(1+z)^3\Bigl(\Omega_{M,0}(z)+\Sigma_M(z)\Bigr)+(1+z)^{3(1+w_m)}\Omega_{m,0}+(1+z)^4\Omega_{rad,0}\Bigr]
\end{equation}
We need to evaluate the entropic terms. Using the relations
\begin{equation}
\frac{kT}{V}=\frac{kT(t_0)}{V(t_0)}\Bigl(\frac{a(t_0)}{a(t')}\Bigr)^4=\rho_T(t_0)\Bigl(\frac{a(t_0)}{a(t')}\Bigr)^4
\end{equation}
and $\Omega_{T,0}=\frac{8\pi G}{3c^2H_0^2}\rho_T(t_0)$ we can write the entropic integrals $\Sigma_k=\Omega_{T,0}J_k$, $k=0,M$ where
\begin{equation}
J_k(t,t_0) = \int \limits_{t_0}^t \Bigl(\frac{a(t')}{a(t_0)}\Bigr)^{3(1+w_k)-4}\frac{dS_k}{dt'} dt'
\end{equation}
The entropies $S_0(t)+S_M(t)=S'=const$ so that $\frac{dS_M}{dt}=-\frac{dS_0}{dt}$. To calculate the entropic integrals $J_0(z,0)$ and $J_M(z,0)$ we shall make use of the linear approximation of Taylor expansion of $\frac{dS_0}{dt}$. The entropy $S_0(t)$ is a functional $S_0(\Sigma_\lambda(t))$ of the entropy $\Sigma_\lambda(t)$ of the space quanta of Dark Energy. Then the part of the integrand in (6.5)
\begin{equation}
\frac{dS_0}{dt'}dt'= \frac{dS_0}{d\Sigma_\lambda}\frac{d\Sigma_\lambda}{d\bar{a}}d\bar{a}
\end{equation}
where $\bar{a}=a(t)/a(t_\omega)$. With $\Sigma_\lambda$ given by (5.3) we define $\tilde{\Sigma}_\lambda=\Sigma_\lambda-2$. In the approximation of  small $\bar{a}$ we find
\begin{eqnarray}
\tilde{\Sigma}_\lambda & = & \frac{1}{\ln 2}
\Bigl(-\frac{1}{2}\bar{a}^4 + 2\bar{a}^2 \ln \bar{a} - \bar{a}^2 \ln 2 \Bigr)\\
\nonumber
\frac{d\Sigma_\lambda}{d\bar{a}} & = & \frac{2}{\ln 2}
\Bigl(-\bar{a}^3+ 2\bar{a} \ln \bar{a} + \bar{a}(1-\ln2) \Bigr)
\end{eqnarray}
From the Taylor expansion for $S_0(\Sigma_\lambda)$ at $\Sigma_{\lambda,0}$ we find in a linear approximation 
\begin{equation}
\frac{dS_0}{d\Sigma_\lambda} = 
\frac{dS_0(\Sigma_{\lambda,0})}{d\Sigma_\lambda} + 
\frac{d^2S_0(\Sigma_{\lambda,0})}{d\Sigma_\lambda^2} \bigl(\tilde{\Sigma}_\lambda-\tilde{\Sigma}_{\lambda,0} \bigr)=C_0+C_1\tilde{\Sigma}_\lambda
\end{equation}
The entropic integrals then read
\begin{eqnarray}
J_0(z,0)= {\bar{a}_0}^4\int \limits_{\bar{a}_0}^{\bar{a}} 
\Bigl(\frac{1}{\bar{a}'}\Bigr)^4 \Bigl[C_0+C_1\tilde{\Sigma}_\lambda\Bigr]\frac{d\Sigma_\lambda}{d\bar{a}}  
d\bar{a}=J_{0A}+J_{0B}\\
J_M(z,0)=\bar{a}_0\int \limits_{\bar{a}_0}^{\bar{a}} \frac{1}{\bar{a}'} \bigl[C_0+C_1\tilde{\Sigma}_\lambda\Bigr]\frac{d\Sigma_\lambda}{d\bar{a}}  
d\bar{a}=J_{MA}+J_{MB}
\end{eqnarray}
With the definitions of entropic parameters $A_T=\Omega_{T,0}2C_0/\ln2$ and $B_T= \Omega_{T,0}2C_1/(\ln2)^2$ we write for $k=0,M$
\begin{equation}
\Omega_{T,0}J_{k,A}=A_T\Sigma_{k,A}, \quad
\Omega_{T,0}J_{k,B}=B_T\Sigma_{k,B}
\end{equation}
Then the Entropic Models E are defined by the expressions for $\tilde{\Omega}_0(z)$ and $\tilde{\Omega}_{M}(z)$
\begin{eqnarray}
\tilde{\Omega}_0(z) & = & \Omega_{0,0}+A_T\Sigma_{0,A}+B_T\Sigma_{0,B}=
\Omega_{0,0}+\Sigma_{0}\\
\tilde{\Omega}_{M}(z) & = & (1+z)^3\Bigl[\Omega_{M,0}+A_T\Sigma_{M,A}+B_T\Sigma_{M,B}\Bigr]=
(1+z)^3\Bigl[\Omega_{M,0}+\Sigma_{M}\Bigr]
\end{eqnarray}
Non-Entropic Models L are defined by $A_T=B_T=0$. 

Neglecting the first terms $\bar{a}^4$ and $\bar{a}^3$ in the equations (6.7) the explicit forms of the entropic integrals $\Sigma_{k,A}$ and $\Sigma_{k,B}$, $k=0,M$ read 
\begin{eqnarray}
\Sigma_{0A} & = & -((1+z)^2-1)\frac{2-\ln2+2\ln \bar{a}_0}{2}\bar{a}_0^2
                  +(1+z)^2(\ln (1+z))\bar{a}_0^2\\
\Sigma_{0B} & = & -\frac{4}{3}(\ln (1+z))^3\bar{a}_0^4 +(\ln (1+z))^2 
                   (1-2\ln2+4\ln\bar{a}_0)\bar{a}_0^4\\
\nonumber                    
            &   & -\ln (1+z)\Bigl[2\ln\bar{a}_0\bigl(1-2\ln2+
                   2\ln\bar{a}_0 \bigr)-\ln 2(1-\ln 2)\Bigr]\bar{a}_0^4\\
\Sigma_{MA} & = & +\frac{\ln(1+z)}{1+z}2\bar{a}_0^2 
                  -\frac{z}{1+z}\bar{a}_0^2(1+\ln2-2\ln\bar{a}_0)\\
\Sigma_{MB} & = & -\frac{(\ln(1+z))^2}{(1+z)^3}\frac{4 \bar{a}_0^4}{3}
                  +\frac{\ln(1+z)}{(1+z)^3}\bar{a}_0^4
    \Bigl[\frac{8\ln\bar{a}_0}{3}-\frac{2}{9}-\frac{4 \ln 2}{3}\Bigr]\\
\nonumber
            &   & -\Bigl[\frac{1}{(1+z)^3}-1.0\Bigr]\bar{a}_0^4
     \Bigl[\frac{4}{3}(\ln \bar{a}_0)^2-\frac{2}{3} \ln(\bar{a}_0)\bigl(\frac{1}{3}+2\ln2\bigr)
     +\frac{2}{27}+\frac{\ln2}{9}+\frac{(\ln2)^2}{3}\Bigr]
\end{eqnarray}
We have verified numerically that these integrals satisfy the condition from the definitions (6.5)
\begin{equation}
\frac{dJ_0}{dz}+(1+z)^3\frac{dJ_M}{dz}=0
\end{equation}

Finally, an important note on our notations. For $w_m=0$ $\Omega_M$ has the meaning of pure Dark Matter. For $w_m=w_r=-\frac{1}{3}$ $\Omega_M$ means combined Dark Matter and atomic matter $\Omega_{M,m}$ in the extended Hubble function (3.30). 

\section{Kinetic origin of the equations of state of Dark Energy and Dark Matter}

\subsection{Space quanta as non-classical particles}

The expanding and contracting homogeneous and isotropic spacetime of the Cyclic Universe is described by the Robertson-Walker (RW) metric. It is a complex standig gravitational wave with tensor components $h_{ij}(\vec{x},t)=S_{ij}(\vec{x})a^2(t)$ defined in (2.1) and a periodic non-singular scale factor $0<a_{min}\leq a(t) \leq a_{max} <\infty$. In the next Section we show that on cosmological scales Dark Matter and Dark Energy are also complex standing waves described by the same scale factor $a(t)$. We can think of the Cyclic Universe as a periodically expanding and contracting spherical volume of gas of the space quanta at the temperature $T(t)$. The evolution of the Universe is described by the entropic equations that determine the components of the Hubble function. The entropic equations for Dark Energy ($k=0$) and Dark Matter ($k=M)$ can be written in the form  
\begin{equation}
\frac{d\rho_k}{dt}+3H\rho_k=-3Hp_k+\frac{kT}{V}\frac{dS_k}{dt}=
-3Hw_k(t)\rho_k=-3HP_k
\end{equation}
where $P_k$ are effective pressures and $w_k(t)=w_k+\xi_k(t)$ are dynamical equations of state. Here $w_k$ are the entropic equations of state $w_0=-1$ and $w_M=0$, and $\xi_k$ are the dynamical parameters
\begin{equation}
\xi_k(t)=-\frac{1}{3H\rho_k}\frac{kT}{V}\frac{dS_k}{dt}
\end{equation}

We now entertain the possibility that space quanta are non-classical particles that have a mass, relativistic energy and momentum defined in terms of their entropies. What makes the space quanta non-classical particles is their negative momentum which is an allowed solution of the equation $m^2c^4=E^2-|\vec{p}|^2c^2$. With $mc^2=\frac{E}{\gamma(v)}$ this equation reads $|\vec{p}|^2=\frac{E^2}{c^2}\frac{|\vec{v}|^2}{c^2}$. With positive energy $E$ it has two solutions
\begin{eqnarray}
p=+|\vec{p}| & => & p=+\frac{E}{c}\frac{|\vec{v}|}{c}>0, \quad \text{classical particles}\\
\nonumber
p=-|\vec{p}| & => & p=-\frac{E}{c}\frac{|\vec{v}|}{c}<0, \quad \text{non-classical particles}
\end{eqnarray}
Since space quanta are two-qubit quantum states they do not have the definite spin like elementary particles.

We start with the definition of the relative entropy of entanglement~\cite{vedral06}
\begin{equation}
S_E(\rho)=\min \limits_{\sigma \in \mathcal{M}_S}
S(\rho||\sigma) = -S(\rho) + \chi(\rho) \geq0 
\end{equation}
where $\mathcal{M}_S$ is the space of the separable states, $S(\rho||\sigma)= -S(\rho)-Tr(\rho\log \sigma)\geq0$ is the relative entropy, $S(\rho)\geq0$ is the entropy of the state $\rho$ and $\chi(\rho)=-Tr(\rho\log \sigma_\rho)$ where $\sigma_\rho$ is the separable state nearest to the state $\rho$. It follows that
\begin{equation}
S_E(\rho)+S(\rho)=\chi(\rho)\geq0
\end{equation}
Our task is to infer the entropic equations of state $w_0=-1$ for Dark Energy and $w_M=0$ for Dark Matter from these entropies at all times as predicted by the joint dynamics of gravity and thermodynamics in the Section III as well as the dynamical parameters $\xi_0(t)$ and $\xi_M(t)$.

\subsection{Dark Energy}

In the quantum model of Dark Energy we define momentum $p(\rho,t)$, energy $E(\rho,t)$ and the mass $m(\rho,t)$ of the space quantum state $\rho$ by the relations
\begin{eqnarray}
p(\rho,t) & = & -E(\rho,t)/c=-kT(t)\chi(\rho)/c\\
E(\rho,t) & = & +kT(t)\bigl(S_E(\rho)+S(\rho)\bigr)=E(\chi,t)\\
m^2(\rho,t)c^4 & = & E^2(\rho,t)-p^2(\rho,t)c^2
\end{eqnarray}
It follows from (7.8) that the mass of Dark Energy quanta $m(\rho,t)=0$. 
At cosmological scales the space quanta $\rho$ of Dark Energy and Dark Matter do not form an ideal gas because their momentum and energy are time dependent.

To account for the dynamical equations of state from the kinetics of the space quanta we shall assume perturbed radially symmetric number density $N_{DE}(\vec{x},t,\chi)=N_{DE}(r,t,\chi)+\delta N_{DE}(r,t,\chi)$ for Dark Energy and 
$N_{DM}(\vec{x},t,S)=N_{DM}(r,t,S)+\delta N_{DM}(r,t,S)$ for Dark Matter. The total energy content of Dark Energy in the volume $V(t)$ then is given by 
\begin{eqnarray}
E_{DE}(t) & = & \int \limits_{V(t)} dV \int d\chi [N_{DE}(r,t,\chi)+\delta N_{DE}(r,t,\chi)]E(\chi,t)\\
\nonumber
 & = & \int \limits_0^{R(t)} 4\pi r^2 dr<\mathcal{E}_{DE}(r,t)>+
 \int \limits_0^{R(t)} 4\pi r^2 dr 
 <\delta \mathcal{E}_{DE}(r,t)>\\ 
 & = & V(t)\bigl(<\mathcal{E}_{DE}(t)>_V + <\delta \mathcal{E}_{DE}(t)>_V\bigr)
\end{eqnarray} 
where $<\mathcal{E}_{DE}(t)>_V$ is the average energy density and $<\delta \mathcal{E}_{DE}(t)>_V$ the average of its pertutbation. The total pressure $P_{DE}(t)$ exerted by the force $F_A$ of Dark Energy at the surface $A(t)$ of the volume $V(t)$ of radius $R(t)$ is
\begin{eqnarray}
P_{DE}(t) & = & \frac{1}{A(t)}\int \limits_{A(t)}dF_A=
\frac{1}{A(t)}\int \limits_{A(t)}dA\frac{dF_A}{dA}\\
\nonumber
  & = & \frac{1}{A(t)}\int \limits_{A(t)}dA\int d\chi
  [N_{DE}(R(t),t,\chi)+\delta N_{DE}(R(t),t,\chi)]p(\chi,t)c\\
\nonumber  
  & = & \frac{1}{A(t)}\int \limits_{A(t)}dA\int d\chi 
  [N_{DE}(R(t),t,\chi)+\delta N_{DE}(R(t),t,\chi)][-E(\chi,t)/c]c\\  
  & = & -<\mathcal{E}_{DE}(t)>_A-<\delta \mathcal{E}_{DE}(t)>_A
\end{eqnarray}
The energy density of Dark Energy must be the same for its volume and at its surface so that $<\mathcal{E}_{DE}(t)>_V=<\mathcal{E}_{DE}(t)>_A$. Substituting $<\mathcal{E}_{DE}(t)>_V$ from (7.10) into (7.12) we find
\begin{equation}
P_{DE}(t) = (-1+\xi_0)\frac{E_{DE}(t)}{V(t)}=(-1+\xi_0)\rho_0(t)
\end{equation}
This shows that the entropic equation of state of Dark Energy is $w_0=-1$ at all times in agreement with the $\Lambda$CDM Model and Section III. The dynamical parameter is
\begin{equation}
\xi_0(t)=\bigl(<\delta \mathcal{E}_{DE}(t)>_V-<\delta \mathcal{E}_{DE}(t)>_A\bigr)\frac{1}{\rho_0(t)}=-\frac{1}{3H\rho_0}\frac{kT}{V}\frac{dS_0}{dt}
\end{equation}

\subsection{Dark Matter}

In the quantum model of Dark Matter space quanta carry no entanglement so that $S_E(\rho)=0$. We define their momentum $p(\rho,t)$, energy $E(\rho,t)$ and the mass $m(\rho,t)$ by the relations
\begin{eqnarray}
p(\rho,t) & = & -E(\rho,t)\frac{v(\rho,t)}{c^2} = 0+kT(t)\delta S(\rho,t)/c\\
E(\rho,t) & = & +kT(t)S(\rho)=E(S,t)\\
m^2(\rho,t)c^4 & = & E^2(\rho,t)-p^2(\rho,t)c^2
\end{eqnarray} 
On cosmological scales Dark Matter quanta carry no momentum but admit its fluctuation due to the fluctuation of the entropy $\delta p=kT(t)\delta S(\rho,t)/c$ where $\delta S(\rho,t)$ is some fraction of $S(\rho)$. We require a negative momentum $p<0$ and set $\delta S(\rho,t)=-\kappa(t)S$ where the fraction $\kappa(t)>0$. The positivity condition $m^2(\rho,t)c^4=(kT(t)S)^2(1-\kappa^2)>0$ implies that $0<\kappa(t)<1$.

Using the relativistic relation $Ev=|p|c^2$ we can formally assign the "speed" $v=c$ to the quanta of Dark Energy and a "speed" 
\begin{equation} 
v(\rho,t)=\frac{|\delta S(\rho,t)|}{S(\rho)}c= \kappa(t)c
\end{equation}
to the quanta of Dark Matter. Space quanta are not matter particles that move in Space. They are the Space. We cannot interpret the quantity $v(\rho,t)$ as the rate of change of position of the quantum state $\rho$. However we can view   
it as a relation between the energy and momentum of the space quantum $\rho$. Note that these "speeds" do not depend on the states $\rho$. 

For the total energy of Dark Matter we obtain in the first order 
\begin{eqnarray}
E_{DM}(t) & = & \int \limits_{V(t)}dV \int dS 
[N_{DM}(r,t,S)+\delta N_{DM}(r,t,S)]E(S,t)\\
\nonumber
 & = & \int \limits_{V(t)} dV \int dS N_{DM}(r,t,S)E(S,t) +
 \int \limits_{V(t)} dV \int dS\bigl(\delta N_{DM}(r,t,S) E(S,t)\bigr)\\
\nonumber 
 & = & \int \limits_{V(t)} dV 
 [<\mathcal{E}_{DM}(r,t)> + <\delta \mathcal{E}_{DM}(r,t)>]\\
 & = & V(t)\bigl(<\mathcal{E}_{DM}(t)>_V+ <\delta \mathcal{E}_{DM}(t)>_V\bigr)
\end{eqnarray}
With momentum given by $p(\rho,t)=\delta p(S,t)$ the total pressure of Dark Matter is determined by the momentum fluctuations $\delta p(S,t)=\delta E(S,t)/c$. In the first order we obtain
\begin{eqnarray}
P_{DM}(t) & = & \frac{1}{A}\int \limits_{A(t)} dA \int dS
[N_{DM}(R(t),t,S)+\delta N_{DM}(R(t),t,S)][\delta E(S,t)/c]v\\
 & = &  \frac{1}{A}\int \limits_{A(t)} dA <\delta \mathcal{E}_{DM}(R(t),t)>\kappa(t)=
 <\delta \mathcal{E}_{DM}(t)>_A\kappa(t)
\end{eqnarray}
The total pressure then reads
\begin{equation}
P_{DM}(t)=(0+\xi_M(t))\frac{E_{DM}(t)}{V(t)}=(0+\xi_M(t))\rho_M(t)
\end{equation}
This shows that the entropic equation of state of Dark Matter is $w_M=0$ at all times in agreement with the $\Lambda$CDM Model and Section III. The dynamical parameter is
\begin{equation}
\xi_M(t)=<\delta \mathcal{E}_{DM}(t)>_A\kappa(t)\frac{V(t)}{E_{DM}(t)}=<\delta \mathcal{E}_{DM}(t)>_A\kappa(t)\frac{1}{\rho_M(t)}=-\frac{1}{3H\rho_M}\frac{kT}{V}\frac{dS_M}{dt}
\end{equation}

\subsection{"Residual" matter term}

In general, when the energy $\epsilon>0$ of the particles of an ideal gas with Bose-Einstein, Fermi-Dirac or Maxwell-Boltzmann statistics is related to their positive momentum $p>0$ by the relation 
\begin{equation}
\epsilon(p) \approx p^s
\end{equation}
the pressure $P$ and energy density $\frac{E}{V}$ of the gas is given by
\begin{equation}
P=\frac{s}{3}\frac{E}{V}
\end{equation}
independently of the statistics obeyed by the particles~\cite{pathria72}. For $s=1$ we recover the case of radiation.

We now consider the case of static and not necessarily radially symmetric galaxies. In these local systems at small scales momentum $p(\rho,t_0)$ and energy $E(\rho,t_0)$ do not depend on time so we assume (7.25) and (7.26) still apply. The two-qubit quantum states $\rho$ do not have a definite spin but could obey Maxwell-Boltzmann or some more general statistics satisfying (7.25) and (7.26). For classical particles $s=2$ and momentum $p>0$. But space quanta are not classical particles and carry a negative momentum. We can thus consider the equation (7.25) for Dark Energy with $p'=|p(\rho,t_0)|>0$ where $t_0$ is a fixed local time. It follows from (7.6) and (7.7) that $E(\rho,t_0)=p'(\rho,t_0)c$ for all states $\rho$ of Dark Energy so that $s=1$. Then (7.26) reads 
\begin{equation}
P'=|P|=-P=\frac{1}{3}\frac{E}{V}
\end{equation}
where $P'$ and $P$ are the magnitude and the value of the pressure, respectively, with $P=-\frac{1}{3}\frac{E}{V}$. This equation formally implies the relation (5.23) $<p>=-\frac{1}{3}<\rho>$ for the pressure and energy density of galactic Dark Energy with the equation of state $w_r=-\frac{1}{3}$ in agreement with Einstein equations. The probability distributions of Dark Matter, both without and with baryons present, are similarly described by the time independent radial functions of the form $f(r,t_0,S)$ so Einstein equations predict the equation of state $w_r=-\frac{1}{3}$ in both these cases. This necessitates a change in the kinetics of the quantum states of the Dark Matter in the galactic systems. With $\delta S=-\kappa S$ where $0<\kappa<1$, the quanta of Dark Matter acquire a negative momentum $p(\rho,t_0)=kT(t_0)\delta S/c=-\kappa(t_0) kT(t_0)S/c$ and a positive energy $E(\rho,t_0)=kT(t_0)S$. With $p'=|p(\rho,t_0)|$ the equation (7.25) reads
\begin{equation}
E(\rho,t_0)=\frac{c}{\kappa(t_0)}p'(\rho,t_0)
\end{equation}
This implies again $s=1$ leading to (7.27) and thus to $<p>=-\frac{1}{3}<\rho>$ for galactic Dark Matter, in agreement with Einstein equations in the case of  radially symmetric densities $p$ and $q$. . 

We now return to the evolving Universe. The combined pressure of all static and not necessarily radially symmetric galaxies $i$ is
\begin{equation}
P_r(t_0) = \sum \limits_i P_i=-\frac{1}{3}
\sum \limits_i \frac{E_{DE,i}(t_0)+E_{DM,i}(t_0)}{V_i(t_0)}
  = -\frac{1}{3}\sum \limits_i \lambda_i\frac{E_r(t_0)}{V_r(t_0)}
  = -\frac{1}{3}\frac{E_r(t_0)}{V_r(t_0)}
\end{equation}
where $\sum \limits_i \lambda_i=1$ and where $E_r(t_0)$ and $V_r(t_0)$ are the total energy and volume of these galaxies. The evolution of the Universe imposes evolution of the energy density $\rho_r(t)=\frac{E_r(t_0)}{V_r(t_0)}$ described by 
\begin{equation}
\frac{d\rho_r}{dt}+3H\rho_r=-3HP_r=-3Hw_r\rho_r
\end{equation}
with the solution $\rho_r(t)=\Bigl (\frac{a(t_0)}{a(t)}\Bigr)^{3(1+w_r)}\rho_r(t_0)$. We have recovered the "residual" matter term (5.24) in the Hubble function (3.30). The densities $\rho_r(t)$ and $p_r(t)=w_r\rho_r(t)$ describe the whole ensemble of all static (slowly evolving) galaxies. We interpret them as galactic "internal" curvature.

\section{Reconstruction of the Robertson-Walker metric from Dark Energy and Dark Matter}

Applied to the cosmological scales the full Einstein equation reads
\begin{equation}
G_{\mu \nu} =\frac{8 \pi G}{c^2}T_{\mu \nu}=\frac{8 \pi G}{c^2}
\Bigl(T^\Lambda_{\mu \nu} +\sum \limits_{k=0}^{rad} T^k_{\mu \nu}\Bigr)
\end{equation}
where the cosmological constant term $T^\Lambda_{\mu \nu}=-\rho_\Lambda g_{\mu \nu}$ and the "material" terms $ T^k_{\mu \nu}$ are given by (5.13). The solution of this equation is expected to be the Robertson-Walker metric $g^{RW}_{\mu \nu}=g^{(0)}_{\mu \nu}
+g^{(c)}_{\mu \nu}$ where the flat and curvature terms are given by~\cite{weinberg08}
\begin{eqnarray}
g^{(0)}_{\mu \nu} & = & \text{diag} (-1,a^2(t),a^2(t),a^2(t))\\
\nonumber
g^{(c)}_{ij} & = & a^2(t)k\frac{r^2\Omega_i \Omega_j}{R_0^2-kr^2}, \quad g^{(c)}_{00}=g^{(c)}_{i0}=g^{(c)}_{0j}=0
\end{eqnarray}
where $\Omega_i=\frac{x^i}{r}$, $i=1,3$ are defined following the equation (5.51). The Robertson-Walker metric is a gravitational field that describes the homogeneous and isotropic Space at large cosmological scales. The gravitational fields of Dark Energy and Dark Matter at these scales arise from the quantum structure of the Space. Since they describe the same Space together they must be equal to the Robertson-Walker metric. 

With the presence of the homogeneous Baryonic matter in the Universe the gravitational fields of Dark Energy and Dark Matter are given by 
\begin{eqnarray}
g_{\mu \nu}^{MB} & = & g^{(0)}_{\mu \nu}+h_{\mu \nu}^{MB}
=g^{(0)}_{\mu \nu}+h_{\mu \nu}^{MV}\Bigl(1+4\pi\chi_M \Bigr)\\
g_{\mu \nu}^{EB} & = & g^{(0)}_{\mu \nu}+h_{\mu \nu}^{EB}
=g^{(0)}_{\mu \nu}+h_{\mu \nu}^{EV}\Bigl(1+4\pi\chi_E \Bigr)
\end{eqnarray}
The combined field is a superposition of the two fields that equals to the Robertson-Walker metric   
\begin{equation}
g^{RW}_{\mu \nu}=\cos^2 \eta g_{\mu \nu}^{MB} + \sin^2 \eta g_{\mu \nu}^{EB}
\end{equation}
The Baryonic gravitational field is taken into account in $g_{\mu \nu}^{MB}$ and $g_{\mu \nu}^{EB}$. 

There are two equivalent forms for $h_{\mu \nu}^{MB}$ and $h_{\mu \nu}^{EB}$. With the displaced static probability distribution $p(\vec{x},S) \to p'(\vec{x}',S)$ where $x_\mu'=x_\mu'(\vec{x})$ we find
\begin{eqnarray}
h^{MB}_{\mu \nu}(\vec{x}) & = & (1+4 \pi \chi_M(\vec{x})) h^{MV}_{\mu \nu}(\vec{x})=
(1+4 \pi \chi_M)a^{n_{\mu\nu}}r_0^2 \int \limits_0^2 dS \frac{1}{p(\vec{x},S)}
\frac{\partial p(\vec{x},S)}{\partial x^\mu}\frac{\partial p(\vec{x},S)}{\partial x^\nu}\\
  & = & 
 a^{n_{\mu\nu}}r_0^2 \int \limits_0^2 dS \frac{1}{p'(\vec{x}',S)}
\frac{\partial p'(\vec{x}',S)}{\partial x^\mu}\frac{\partial p'(\vec{x}',S)}{\partial x^\nu}\\
  & = & 
a^{n_{\mu\nu}}r_0^2 \sum \limits_{k,m=1}^3 \Biggl[\int \limits_0^2 dS \frac{1}{p'(\vec{x}',S)}
\frac{\partial p'(\vec{x}',S)}{\partial x^{'k}}\frac{\partial p'(\vec{x}',S)}{\partial x^{'m}}\Biggr] \Biggl(\frac{\partial x^{'k}}{\partial x^\mu}
\frac{\partial x^{'m}}{\partial x^\nu}\Biggr)\\
   & = &
a^{n_{\mu\nu}}r_0^2 \sum \limits_{k,m=1}^3 \Biggl[\int \limits_0^2 dS 
\frac{1}{p'(\vec{x}',S)}
\frac{\partial p'(\vec{x}',S)}{\partial x^{'k}}\frac{\partial p'(\vec{x}',S)}{\partial x^{'m}}\Biggr] \Biggl(\frac{\partial x^{'k}}{\partial r}
\frac{\partial x^{'m}}{\partial r}\Biggr)\Omega_\mu \Omega_\nu
\end{eqnarray}
where $n_{00}=0$, $n_{i0}=n_{0j}=1$, $n_{ij}=2$ and $\Omega_0=0$. With the displaced static probability distribution 
$q(\vec{x},\chi) \to q'(\vec{x}',\chi)$ we can write $h_{\mu \nu}^{EB}$ in a similar form.

Since $g^{(c)}_{00}=0$ it follows from (8.6) and (8.7) that $p$ and $p'$ (and similarly $q$ and $q'$) do not depend on time $t$. The consistency of the expressions (8.6),(8.7) and (8.8), (8.9) for the spatial components $ij$ then implies time independence of $\chi_M$ (and similarly $\chi_E$). A comparison with  $g^{(c)}_{ij}$ then implies that $\eta$ does not depend on time but can depend on $r$. It follows fom (8.9) that
the terms multiplying the angular factors $\Omega_i\Omega_j$ can be negative as required by (8.12) and (8.13) below. Assuming $\frac{\partial x^{'k}}{\partial x^\mu}=const\delta_{x^{'k},x^\mu}$ on galacic scales we recover the relations (5.20) and (5.21).

Following (5.9) and (5.12) and assuming (5.48) and (5.55) we can write at large scales
\begin{eqnarray}
h_{\mu \nu}^{MV} & = & a^{n_{\mu\nu}}r^2_0\frac{2}{\alpha^2} \frac{\partial \alpha}{\partial x^\mu}\frac{\partial \alpha}{\partial x^\nu}
=a^{n_{\mu\nu}}\Bigl(r_0\frac{\sqrt{2}}{\alpha}
\frac{d \alpha}{dr}\Bigr)^2 \Omega_\mu \Omega_\nu\\
h_{\mu \nu}^{EV} & = & a^{n_{\mu\nu}}r^2_0\frac{\chi_m}{\beta^2} \frac{\partial \beta}{\partial x^\mu}\frac{\partial \beta}{\partial x^\nu} 
=a^{n_{\mu\nu}}\Bigl(r_0\frac{\sqrt{\chi_m}}{\beta}
\frac{d \beta}{dr}\Bigr)^2 \Omega_\mu \Omega_\nu
\end{eqnarray}
In (8.10) and (8.11) we assume static and radially symmetric $\alpha(r,t)$ and $\beta(r,t)$, respectively, and define $\Omega_0=0$. Note that at large scales not necessarily $|h_{\mu \nu}^{EV}| \ll 1$ and $|h_{\mu \nu}^{MV}| \ll 1$ (see below). After some cancellations the condition (8.5) implies
\begin{eqnarray}
\frac{kr^2}{R_0^2-kr^2} & = & \cos^2 \eta
\Bigl(r_0\frac{\sqrt{2}}{\alpha}\frac{d \alpha}{dr}\Bigr)^2
\Bigl(1+4\pi\chi_M \Bigr)\\
\nonumber
  & + & \sin^2 \eta 
\Bigl(r_0\frac{\sqrt{\chi_m}}{\beta}\frac{d \beta}{dr}\Bigr)^2
\Bigl(1+4\pi\chi_E \Bigr)
\end{eqnarray}
We seek the solution of (8.12) for $k=-1$ in the form
\begin{eqnarray}
k=-1 & = & \cos^2 \eta \Bigl(1+4\pi\chi_M \Bigr)=
\sin^2 \eta \Bigl(1+4\pi\chi_E \Bigr)\\
  &  & \Bigl(r_0\frac{\sqrt{2}}{\alpha}\frac{d \alpha}{dr}\Bigr)^2=
\cos^2 \xi \frac{r^2}{R_0^2+r^2} \\
  &  & \Bigl(r_0\frac{\sqrt{\chi_m}}{\beta}\frac{d \beta}{dr}\Bigr)^2=
\sin^2 \xi \frac{r^2}{R_0^2+r^2}
\end{eqnarray}
The equations (8.13) imply
\begin{equation}
\frac{1}{1+4\pi\chi_M}+\frac{1}{1+4\pi\chi_E}=-1
\end{equation}
For $|h^{MB}_{\nu}| \ll 1$ we have $r_0 \frac{\sqrt{2}}{\alpha}\frac{d \alpha}{dr} \ll 1$ which implies $\frac{r^2}{R_0^2+r^2}\ll 1$. Since all functions are analytical we can analytically continue the equation (8.14) to $r\to \infty$. The same proceedure applies to (8.15). Integrating (8.14) and (8.15) from $r=0$ to $r$ we find the solutions of the equations (8.14) and (8.15) read
\begin{eqnarray}
\alpha(r) & = & \alpha(0)\exp \Biggl[\int \limits_0^r 
\sqrt{\frac{\cos^2 \xi}{2r_0^2}\frac{r^2}{R_0^2+r^2}}dr \Biggr]\\
\nonumber
\beta(r) & = & \beta(0)\exp \Biggl[\int \limits_0^r 
\sqrt{\frac{\sin^2 \xi}{\chi_m r_0^2}\frac{r^2}{R_0^2+r^2}}dr \Biggr]
\end{eqnarray}
The gravitational fields (8.3) and (8.4) of Dark Matter and Dark Energy now have a new equivalent form with $k=-1$
\begin{eqnarray}
g^{MB}_{\mu \nu} & = & g^{(0)}_{\mu \nu}+a^{n_{\mu \nu}}\frac{\cos^2 \xi}{\cos^2 \eta} \frac{kr^2}{R_0^2-kr^2} \Omega_\mu \Omega_{\nu}\\
g^{EB}_{\mu \nu} & = & g^{(0)}_{\mu \nu}+a^{n_{\mu \nu}}\frac{\sin^2 \xi}{\sin^2 \eta} \frac{kr^2}{R_0^2-kr^2} \Omega_\mu \Omega_{\nu}
\end{eqnarray}

In general the parameter $\xi$ is a function of $r$. In a notable special case where $\xi(r)\equiv \eta(r)$ the gravitational fields of Dark Energy and Dark Matter both coincide with the Robertson-Walker metric
\begin{equation}
g_{\mu \nu}^{EB}=g_{\mu \nu}^{MB}=g_{\mu \nu}^{RW}
\end{equation}
This possibility is physically very intriguing. For $\xi(r) \approx 0$ the gravitational field of Dark Energy is simply the flat component of the Robertson-Walker metric $g^{EB}_{\mu\nu} \approx 
g^{(0)}_{\mu \nu}$. 

At the scales of galaxies $(G)$ and clusters $(C)$ as well as the voids the Baryonic field $h_{\alpha,\beta}^{B,GC} \neq h_{\alpha,\beta}^{Bm}$ so that $h_{\mu \nu}^{MB,GC} \neq h_{\mu \nu}^{MB,RW}$. The local gravitational field of the Space is then a superposition of the local fields 
$g_{\mu \nu}^{MB,GC}=g_{\mu \nu}^{(0)}+h_{\mu \nu}^{MB,GC}$ and 
$g_{\mu \nu}^{EB,GC}=g_{\mu \nu}^{(0)}+h_{\mu \nu}^{EB,GC}$ 
\begin{equation}
g_{\mu \nu}^{GC}
=g_{\mu \nu}^{(0)}+\cos^2 \eta^{GC} h_{\mu \nu}^{MB,GC} 
+ \sin^2 \eta^{GC} h_{\mu \nu}^{EB,GC}=
g_{\mu \nu}^{(0)}+h_{\mu \nu}^{GC}
\end{equation}
For $|h_{\mu \nu}^{GC}|\ll 1$ we reset $g_{\mu \nu}^{(0)}=\eta_{\mu \nu}$ so that
\begin{equation}
g_{\mu \nu}^{GC} = \eta_{\mu \nu} + h_{\mu \nu}^{GC}
\end{equation}

\section{Recovery of Friedmann equations from the gravitational fields of Dark Energy and Dark Matter.}

We can view the Robertson-Walker metric as a solution of the Einstein equations for a gravitational field sourced by the energy-momentum stress tensor of the cosmic fluid with energy density $\rho$ and pressure $p$. Einstein equations are then reduced to the Friedmann equations (2.2) and (2.3). Similarly, we can view the gravitational field of Dark Energy $g_{\mu \nu}^{EB}$ (Dark Matter $g_{\mu \nu}^{MB}$) at cosmological scales as the solution of Einstein equations sourced by an effective energy-momentum stress tensor of Dark Energy (Dark Matter) formed by the fractions $\bar{\rho}_0$ and $\bar{p}_0$ ($\bar{\rho}_M$ and $\bar{p}_M$) of the energy density $\rho$ and pressure $p$ with
$\bar{\rho}_0=f\rho$, $\bar{p}_0=fp$ and $\bar{\rho}_M=g\rho$, $\bar{p}_M=gp$ where $\sin^2 \eta(0) f + \cos^2 \eta(0) g=1$.

Following Weinberg~\cite{weinberg08} we write Einstein equations with no cosmological constant in the form
\begin{equation}
R_{\mu \nu}=-\frac{8\pi G}{c^4} S_{\mu \nu}
\end{equation}
where $R_{\mu \nu}$ is the Ricci tensor and $S_{\mu \nu}$ is given in terms of the energy-momentum stress tensor
\begin{eqnarray}
S_{\mu \nu} & = & T_{\mu \nu}-\frac{1}{2}g_{\mu \nu} T^\mu_\mu\\
T_{00} & = & \rho, \quad T_{i0}=0, \quad T_{ij}=a^2 p g_{ij}
\end{eqnarray}
For the Robertson-Walker metric with $k=-1$ we define
\begin{equation}
g_{ij}^{RW} = a(t)^2\Bigl(\delta_{ij}-\frac{x^i x^j}{R_0^2+r^2}\Bigr)=
a(t)^2 \tilde{g}_{ij}
\end{equation}
The Robertson-Walker field is a superposition of the Dark Matter and Dark Energy fields with $k=-1$ for which we write 
\begin{eqnarray}
g_{ij}^{MB} & = & a(t)^2\Bigl(\delta_{ij}-C(r)\frac{x^i x^j}{R_0^2+r^2}\Bigr)=
a(t)^2 \bar{g}_{ij}^M\\
g_{ij}^{EB} & = & a(t)^2\Bigl(\delta_{ij}-S(r)\frac{x^i x^j}{R_0^2+r^2}\Bigr)=
a(t)^2 \bar{g}_{ij}^E
\end{eqnarray}
where $C(r)=\frac{\cos^2 \xi}{\cos^2 \eta}$ and $S(r)=\frac{\sin^2 \xi}{\sin^2 \eta}$. For the Einstein equations of Dark Matter we find
\begin{eqnarray}
R_{ij}^Mc^2 & = & \tilde{R}_{ij}^Mc^2 - 2\dot{a}^2\bar{g}_{ij}^M -
a\ddot{a} \bar{g}_{ij}^M =-\frac{4\pi G}{c^2} (\bar{\rho}_M - \bar{p}_M)a^2 \bar{g}_{ij}^M\\
R_{00}^Mc^2 & = & 3\frac{\ddot{a}}{a}=
-\frac{4\pi G}{c^2} (3\bar{p}_M+\bar{\rho}_M)\\
R_{i0}^M & = & 0, \quad R_{0i}^M=0
\end{eqnarray}
$\bar{\rho}_M$ and $\bar{p}_M$ are the effective energy density and pressure of Dark Matter. The purely spatial term $\tilde{R}_{ij}^M$ was calculated and it has the form
\begin{equation}
\tilde{R}_{ij}^M = A^M(r)\delta_{ij}+B^M(r)\tilde{g}_{ij}
\end{equation}
where the terms $A^M(r)$ and $B^M(r)$ depend on $C(r)$ and $\frac{dC}{dr}$. We are interested in the point $\vec{x}=0$ where the observer is located and makes his observations and measurements. At $r=0$ the spatial Ricci tensor $\tilde{R}_{ij}^M=\frac{2C(0)}{R_0^2}\delta_{ij}$ so that it can be rewritten in the form
\begin{equation}
\tilde{R}_{ij}^M=\frac{2C(0)}{R_0^2}\bar{g}_{ij}^M
\end{equation}
This is a relation between two three-tensors which must be true in all coordinate systems related by transformations of the point $\vec{x}=0$ into another point in space. The equation (9.7) then implies
\begin{equation}
\frac{2C(0)c^2}{R_0^2a^2}-\frac{2\dot{a}^2}{a^2}-\frac{\ddot{a}}{a}=
-\frac{4\pi G}{c^2} (\bar{\rho}_M-\bar{p}_M)
\end{equation}
Using (9.8) to eliminate the $\ddot{a}$ term in (9.12) these two equations take the form
\begin{eqnarray}
\bar{\rho}_M+C(0)\rho_c & = & \frac{3c^2}{8\pi G}H^2\\
\bar{p}_M + C(0)p_c & = & \frac{3c^2}{8\pi G}\bigl(-H^2-\frac{2}{3}\frac{dH}{dt} \bigr)
\end{eqnarray}
where the curvarure density and pressure $\rho_c$ and $p_c$ are defined by (2.4). 

For the Einstein equations of Dark Energy we find
\begin{equation}
R_{ij}^Ec^2  =  \tilde{R}_{ij}^Ec^2 - 2\dot{a}^2\bar{g}_{ij}^E -
a\ddot{a} \bar{g}_{ij}^E =
-\frac{4\pi G}{c^2} (\bar{\rho}_0 - \bar{p}_0)a^2 \bar{g}_{ij}^E
\end{equation}
The Ricci tensor $\tilde{R}_{ij}^E$ has a form identical to (9.10) with $C(r)$ replaced by $S(r)$. At $r=0$ it reads $\tilde{R}_{ij}^E=\frac{2S(0)}{R_0^2}\delta_{ij}$ so that again we can write $\tilde{R}_{ij}^E=\frac{2S(0)}{R_0^2}\bar{g}_{ij}^E$. This leads to equations similar to (9.13) and (9.14) with $\bar{\rho}_0$ and $\bar{p}_0$ being the effective energy density and pressure of Dark Energy
\begin{eqnarray}
\bar{\rho}_0+S(0)\rho_c & = & \frac{3c^2}{8\pi G}H^2\\
\bar{p}_0 + S(0)p_c & = & \frac{3c^2}{8\pi G}\bigl(-H^2-\frac{2}{3}\frac{dH}{dt} \bigr)
\end{eqnarray}
Next we multiply the equations (9.13) and (9.14) by $\cos^2 \eta(0)$ and the equations (9.16) and (9.17) by $\sin^2 \eta(0)$. Adding such equations (9.13) and (9.16) and similarly adding (9.14) and (9.17) we recover the Friedmann equations (2.2) and (2.3) in the form reminiscent of (8.5) 
\begin{eqnarray}
\cos^2 \eta(0)\bar{\rho}_M+\sin^2 \eta(0)\bar{\rho}_0 +\rho_c & = & \rho+\rho_c =
\frac{3c^2}{8\pi G}H^2\\
\cos^2 \eta(0)\bar{p}_M+\sin^2 \eta(0)\bar{p}_0 +p_c & = & p+p_c =
\frac{3c^2}{8\pi G}\bigl(-H^2-\frac{2}{3}\frac{dH}{dt} \bigr)
\end{eqnarray}
Following the Sections III.A-III.C we recover the General form (3.18) of the Hubble function $H^2(z)$ and following the Section IV its Entropic form (5.30).
We have connected the gravitational fields (8.18) and (8.19) of Dark Matter and Dark Energy at cosmological scales to the measurable Hubble function with the relations (9.13)-(9.14) and (9.16)-(9.17), respectively, involving the fields parameters $C(0)$ and $S(0)$ and the parameters of their energy-momentum stress tensors.

Finally we note that we can determine the fractions $g$ and $f$ using (9.13) and (9.16). Assuming $\rho_\Lambda=0$ in the Friedmann equation (2.2) we can write
\begin{eqnarray}
\bar{\rho}_M & = & g\Big(\frac{3c^2}{8\pi G}H^2-\rho_c\Bigr)=
\frac{3c^2}{8\pi G}H^2-C(0)\rho_c\\
\nonumber
\bar{\rho}_0 & = & f\Bigl(\frac{3c^2}{8\pi G}H^2-\rho_c\Bigr)=
\frac{3c^2}{8\pi G}H^2-S(0)\rho_c
\end{eqnarray}
Hence we obtain
\begin{equation}
g= \frac{\frac{3c^2}{8\pi G}H^2-C(0)\rho_c}{\frac{3c^2}{8\pi G}H^2-\rho_c},
\quad
f= \frac{\frac{3c^2}{8\pi G}H^2-S(0)\rho_c}{\frac{3c^2}{8\pi G}H^2-\rho_c}
\end{equation}

\section{Conclusions and Outlook}

We have constructed a cosmology of Cyclic Universe governed by joint laws of the General theory of relativity, Thermodynamics and Quantum Information theory. Einstein's theory of gravity and Thermodynamics contribute a general form of the Hubble function which predicts Dark Energy and Dark Matter as well as the acceleration-deceleration transitions of the Universe expansion. Thermodynamics also predicts, at all $z$, the equation of state of Dark Energy $w_0=-1$ and Dark Matter $w_M=0$. Quantum Information theory elucidates the physical nature of the Dark Energy and Dark Matter in terms of the quantum structure of the spacetime. Thermodynamics of the components of the Universe and the quantum structure of space specify our Entropic Model E. 

The central tenets of the Entropic Model are three ideas: (i) Space is gravitational fields (ii) Space has a quantum structure (iii) Dark Energy and Dark Matter are the Space. We identify the non-local entangled space quanta with Dark Energy and the non-entangled space quanta with Dark Matter. Dark Energy and Dark Matter are described by probability distribution functions $p(\vec{x},t,S)$ and $q(\vec{x},t,\chi)$ of entropies of their quanta. The gravitational fields of Dark Energy and Dark Matter are given by two quantum duality relations in terms of the Fisher metric defined by the distributions $p$ and $q$, respectively. Baryonic matter perturbs these distributions. The perturbations $\delta p$ and $\delta q$ are related to the Baryonic gravitational field by Baryonic quantum duality relations. 

This quantum model of the spacetime determines the entropic (dynamical) terms $\Sigma_0(t)$ and $\Sigma_M(t)$ of Dark Energy and Dark Matter, respectively, in terms of the entropy $\Sigma_\lambda(t)$ of the average quantum state $\rho_\lambda(t)$ of the Dark Energy. The theory also predicts a new cosmological component $\rho_r(t)$ with equation of state $w_r=-\frac{1}{3}$. Called "residual" matter term it plays the role of an "internal" curvature term. The consistency with the "external" curvature $\rho_c(t)$ of the Robertson-Walker metric demands that $k=-1$. The quantum structure of space thus predicts that our spacetime is anti-de Sitter spacetime. The prediction of positive spatial curvature density $\Omega_{c,0}$ imposes separate constraints on the cosmological parameters $\Omega_{Mm,0}$ and $\Omega_{0,0}$. 

To prove the consistency of the quantum model of the spacetime with the thermodynamical extension of the Friedmann equations presented in the Section III we have used the kinetics of space quanta to derive the entropic equations of state $w_0=-1$ and $w_M=0$ for Dark Energy and Dark Matter, respectively, as well as the corresponding dynamical parameters $\xi_0(z)$ and $\xi_M(z)$. Notable achievement of this model is the demonstration in the Section VII.D that the "residual" matter term in the Hubble function arises from all slowly evolving ("static") galaxies in agreement with the predictions of the Einstein equations in the case of radially symmetric densities $p$ and $q$. 

Notwithstanding the interpretation of this new term as an "internal" curvature we treat the "residual" matter as a component of the Universe with a particle structure and subject to the Euler equations. The densities $\rho_r(t)$ and $p_r(t)=w_r\rho_r(t)$ describe the entire ensemble of the $N_r$ galaxies ("particles") of the Universe now identified as the "residual" matter. Its Euler equations predict the presence of the "residual" matter term in the Hubble function. Its constant entropy $S_r$ means that this system of galaxies as a whole is a closed system subject only to its own cosmic evolution. In particular, galaxies can form evolving clusters, supercluster and the cosmic web. 
 
At cosmological scales Space is described by the Robertson-Walker gravitational field as well as by the gravitational fields of Dark Energy and Dark Matter. Identifying the Robertson-Walker metric with a superposition of the general solutions for the gravitational fields of Dark Energy and Dark Matter we determine the explicit form of these fields in terms of parameters $\xi(r)$ and $\eta(r)$. Assuming effetive energy-momentum stress tensors of Dark Energy and Dark Matter, we use the Einstein equations for these fields to calculate the Friedmann-like equations for each field and show these solutions combine to recover the Friedmann equations proper. This result is a direct consequence of the special values of the spatial Ricci tensors at $\vec{x}=0$ for these two fields. With the parameter $\xi(r)\equiv \eta(r)$ the gravitational fields of Dark Energy and Dark Matter at cosmological scales both coincide with the Robertson-Walker metric. 

In the follow-up paper~\cite{svec17b} we test all our predictions in the fits of the Model E to the Hubble data $H(z)$ and angular diameter distance data $d_A(z)$. The fits to Hubble data confirm the existence of the "residual" matter term with $w_r=-\frac{1}{3}$ and validate the constraints on the cosmological parameters. The fits to $d_A(z)$ determine positive values of $\Omega_{c,0}$ in agreement with the theory. The fits of the Model E to the analytical Model A of the Cyclic Universe developed in a related paper~\cite{svec17a} confirm the equivalence of the two models with an astonishing $\chi^2/dof=0.0000057$. We conclude that Model A and Model E represent the same Cyclic Universe with negative curvature and quantum structure of the spacetime.

The initial energy density and the initial temperature of each cycle of the Cyclic Universe are finite, and there is no indefinite expansion. There are no increases of the entropy of the Cyclic Universe after each cycle that disallowed the early models of Cyclic Universe~\cite{tolman34}. The entropy of the atomic matter increases during the expansion but is fully transformed back into the entropy of radiation during the contraction. This is possible because the atomic matter and radiation form a subsystem connected by the cyclic acale factor $a(t)$ to the larger dynamical system of the whole Universe (see the relations (4.28) and (4.29)). Associated transformations of radiation into atomic matter and vice versa during the evolution arise from their particle interactions and redshifts. The evolution dynamics which conserves the total entropy of the atomic matter and radiation $S''=S_m(t)+S_{rad}(t)=S_m(a(t))+S_{rad}(a(t))$ as well as their total particle numbers $N''=N_m(t)+N_{rad}(t)=N_m(a(t))+N_{rad}(a(t))$ thus allows a Cyclic Cosmology.

In a sequel paper~\cite{svec17c} we use our quantum information model of Dark Energy and Dark Matter to study their gravitational fields in a static and radially symmetric galaxy. We recover the radial acceleration relation of galaxies recently discovered in a study of 153 different galaxies~\cite{mcgaugh16,lelli16} and find that the equation of state of this galactic Dark Matter and Dark Energy system is $w_r=-1/3$. The existence of this "residual" matter term $\rho_r(t)$ in the Hubble function is strongly supported by the agreement of the theory with the observed radial acceleration relation.  

The evidences for the "residual "matter term and anti-de Sitter spatial curvature support our quantum model of the spacetime but the ultimate evidence will be the direct observation of the space quanta in dephasing hadron interactions. First proposed by Hawking in 1982~\cite{hawking82,hawking84} to describe non-unitary interactions of the  particle scattering processes with spacetime metric fluctuations, the dephasing interactions are non-unitary interactions of the produced final hadron states $\rho_f(S)$ with the quantum states $\rho(E)$ of a quantum environment $E$ to form observed hadron states $\rho_f(O)$~\cite{svec13a}. The observed states $\rho_f(O)$ carry information about the diagonal elements of the states $\rho(E)$~\cite{svec13a}. We can identify the quantum environment $E$ with the Space and the quanta $\rho(E)$ with the space quanta. In principle, the measurements of the state $\rho_f(O)$ in suitable hadron interactions could determine average values of the diagonal elements of the local space quanta. 

The theoretical basis of the cyclic/entropic cosmology in the three pillars of Modern Physics renders this cosmology a compelling framework for the analysis of the observations to be made by the ongoing and upcoming astronomical surveys at high redshifts~\cite{DES,LSST,Euclid,WFIRST,spergel15,SKA}.

\acknowledgements

I acknowledge with thanks the technical support of this research by Physics Department, McGill University.


\appendix

\section{Friedmann equation for the scale factor}

The scale factor is not a simple but a complex periodic function of time. Because it has a deep physical meaning we expect this function to be a solution of a fundamental equation for a non-harmonic oscillator. We show that the second Friedmann equation (2.10) is formally equivalent to such an equation. We write (2.10) in the form 
\begin{equation}
\frac{2}{3}\frac{dH}{dt}=-H^2-\frac{8\pi G}{3c^2} \bar{p}
\end{equation}
This equation is a particular form of the Riccati equation~\cite{polyanin02}
\begin{equation}
g(x)\frac{dy}{dx}=f_2(x)y^2+f_1(x)y+f_0(x)
\end{equation}
Using the transform
\begin{equation}
u(x)=\exp\Bigl( -\int \frac{f_2}{g} y dx \Bigr)
\end{equation}
the equation (A2) takes the form
\begin{equation}
f_2g^2\frac{d^2u}{dx^2} + g \Bigl[f_2\frac{dg}{dx}-g\frac{df_2}{dx}-f_1f_2 \Bigr]\frac{du}{dx} +f_0f_2^2 u=0
\end{equation}
In our case $g=\frac{2}{3}$, $f_2=-1$, $f_1=0$, 
$f_0=-\frac{8\pi G}{3c^2}\bar{p}$. With the trasform for $t>t_\alpha$
\begin{equation}
u(t)  =  \exp\Bigl( -\int \limits_{t_\alpha}^{t} \frac{-3}{2} H dt' \Bigr) =  \Bigl(\frac{a(t)}{a(t_\alpha)}\Bigr)^\frac{3}{2}
\end{equation}
we find
\begin{equation}
\frac{d^2u}{dt^2} + \frac{3}{2}\Omega^2_p(t) u=0
\end{equation}
This is an equation for simple harmonic motion with time dependent angular frequency $\frac{3}{2}\Omega^2_p(t)=\frac{6\pi G}{c^2}\bar{p}$ for the function $u=a^\frac{3}{2}(t)$. The equation for the scale factor itself reads
\begin{equation}
\frac{d^2a}{dt^2} + \Omega^2_p(t)a =
-\frac{1}{2a}\Bigl(\frac{da}{dt}\Bigr)^2
\end{equation}
where $\Omega^2_p(t)=\frac{4\pi G}{c^2}\bar{p}$. This is an equation for  a self-driven non-harmonic oscillator with the time dependent cyclic angular frequency $\Omega^2_p(t)$ that changes signs and the self-driving "force" $-\frac{1}{2a}\Bigl(\frac{da}{dt}\Bigr)^2$. Its solutions are periodic non-harmonic functions $a(t)$ that do not change signs. In a related work~\cite{svec17a} we give a specific example of such sign changing $\Omega^2_p(t)$ and the corresponding non-harmonic periodic scale factor. It is useful to write (A7) in the form
\begin{equation}
\frac{d^2a}{dt^2}+\frac{1}{2a}\Bigl(\frac{da}{dt}\Bigr)^2 + F(a)=0
\end{equation}
where $F(a)=\frac{4\pi G}{c^2}\bar{p}(a)a$ with $\bar{p}$ given by (3.15). With the substitution $y(a)=\frac{da}{dt}$ and $\frac{d^2a}{dt^2}=\frac{dy}{dt}=\frac{dy}{da}\frac{da}{dt}=\frac{dy}{da}y$ the second order equation (A8) becomes a first order equation
\begin{equation}
\frac{dy}{da}=-\frac{1}{2a}y -F(a)\frac{1}{y}
\end{equation}
This is a special case of the equation~\cite{polyanin02}
\begin{equation}
\frac{dy}{dx}=f(x)y^{1+n}+g(x)y+h(x)y^{1-n}
\end{equation}
which can be transformed with a substitution $w=y^n$ into a Riccati equation
\begin{equation}
\frac{dw}{dx}=nf(x)w^2+ng(x)w+nh(x)
\end{equation}
In our case we have $f=0$ and $n=2$ so with $w=y^2$ the equation (A9) becomes
\begin{equation}
\frac{dw}{da}=-\frac{1}{a}w-2F(a)
\end{equation}
This is a first order linear equation $y'(x)+P(x)y=Q(x)$ which has a solution given by (3.3)
\begin{equation}
w(a)=\frac{a_0}{a} \Biggl\{ a(a_0)-2\int \limits_{a_0}^{a} \Bigl(\frac{a'}{a_0}\Bigr) F(a')da' \Biggr\}
\end{equation}
With $w=\Bigl(\frac{da}{dt}\Bigr)^2$ and $H^2=\frac{w}{a^2}$ we recover the solution of the Friedmann equation (3.4), leading to the Hubble function (3.18). 

The equation (3.18) for the predicted Hubble function can be solved for the scale factor
\begin{equation}
a(t)=a(t_0)\pm \Bigl(\frac{da}{dt}\Bigr)_0 \int \limits_{t_0}^t 
\sqrt{\frac{a^2}{a_0^2}\Bigl[\Omega_{0,0}+\Sigma_0(a)\Bigr]+
         \frac{a_0}{a}\Bigl[\Omega_{Mm,0}+\Sigma_{M}(a)\Bigr]+
\frac{a_0^2}{a^2}\Omega_{rad,0}} dt'
\end{equation}
where the $+$ ($-$) sign applies to the expansion (contraction) with $\frac{da}{dt}>0$ ($\frac{da}{dt}<0$). The equation (A14) is a complex integral equation for $a(t)$. The functions $\Sigma_0(a)$ and $\Sigma_{M}(a)$ are unknown. Models of these functions determine the Hubble function $H^2(a)$ in (3.18) but solving the equation (A14) for $a(t)$ may be elusive. Models of $a(t)$ can determine $H^2(a)$ from the definition of the Hubble parameter but strongly restrict models of $\Sigma_0(a)$ and $\Sigma_M(a)$ which must reproduce this $H^2(a)$ in the equation (3.18). We successfully follow this latter approach in our related paper~\cite{svec17a} and in this work.

\section{Two-qubit quantum states}

Two-qubit quantum states $\rho$ are $4\times4$ Hermitian density matrices with trace $Tr\rho(E)=1$. They can be written in two equivalent forms
\begin{eqnarray}
\rho_t & = & \frac{1}{4} \sum \limits_{\mu,\nu=0}^3 t_{\mu\nu} \sigma^\mu \otimes \sigma^\nu\\
\rho_p & = & \sum \limits_{m,n=0}^1 \sum \limits_{m',n'=0}^1 p_{mn,m'n'}|m>|n><m'|<n'|
\end{eqnarray}
where $\rho_t=\rho_p$. The form (B1) is expressed in the Fano basis $\sigma^\mu \otimes \sigma^\nu$ where $\sigma^\mu$ are Pauli matrices. The form (A2) is expressed in the computational basis of two-qubit states 
$|m>|n>$, $m,n=0,1$ where $|0>=|+2>,|1>=|-2>$ represent helicity states of massless gravitons. The computational form follows from the Fano form using Pauli matrices in the single qubit computational basis
\begin{eqnarray}
\begin{array}{rl}
\sigma^1=|0><1|+|1><0|, & \sigma^3=|0><0|-|1><1|\\
i\sigma^2=|0><1|-|1><0|, & \sigma^0=|0><0|+|1><1|
\end{array}
\end{eqnarray}
Using relations (B3) we can relate the matrix elements $p_{mn,m'n'}$ to $t_{\mu\nu}$
\begin{eqnarray}
\begin{array}{rl}
p_{00,00}=\frac{1}{4}\big(t_{00}+t_{30}+t_{03}+t_{33}\bigr), & 
p_{01,01}=\frac{1}{4}\big(t_{00}+t_{30}-t_{03}-t_{33}\bigr)\\
p_{11,11}=\frac{1}{4}\big(t_{00}-t_{30}-t_{03}+t_{33}\bigr), &
p_{10,10}=\frac{1}{4}\big(t_{00}-t_{30}+t_{03}-t_{33}\bigr)
\end{array}
\end{eqnarray}
\begin{eqnarray}
\begin{array}{rl}
p_{00,11}=\frac{1}{4}\big(t_{11}+it_{12}+it_{21}-t_{22}\bigr), &
p_{01,10}=\frac{1}{4}\big(t_{11}-it_{12}+it_{21}+t_{22}\bigr)\\
p_{11,00}=\frac{1}{4}\big(t_{11}-it_{12}-it_{21}-t_{22}\bigr), &
p_{10,01}=\frac{1}{4}\big(t_{11}+it_{12}-it_{21}+t_{22}\bigr)
\end{array}
\end{eqnarray}
\begin{eqnarray}
\begin{array}{rl}
p_{00,10}=\frac{1}{4}\big(t_{10}+t_{13}+i(t_{20}+t_{23})\bigr), &
p_{01,11}=\frac{1}{4}\big(t_{10}-t_{13}+i(t_{20}-t_{23})\bigr)\\
p_{10,00}=\frac{1}{4}\big(t_{10}+t_{13}-i(t_{20}+t_{23})\bigr), &
p_{11,01}=\frac{1}{4}\big(t_{10}-t_{13}-i(t_{20}-t_{23})\bigr)
\end{array}
\end{eqnarray}
\begin{eqnarray}
\begin{array}{rl}
p_{00,01}=\frac{1}{4}\big(t_{01}+t_{31}+i(t_{02}+t_{32})\bigr), &
p_{10,11}=\frac{1}{4}\big(t_{01}-t_{31}+i(t_{02}-t_{32})\bigr)\\
p_{01,00}=\frac{1}{4}\big(t_{01}+t_{31}-i(t_{02}+t_{32})\bigr), &
p_{11,10}=\frac{1}{4}\big(t_{01}-t_{31}-i(t_{02}-t_{32})\bigr)
\end{array}
\end{eqnarray}

The most general form of any separable state $\rho_{sep}$ with dimension 4 is well known state~\cite{bengtsson06,friis16}
\begin{equation}
\rho_{sep}=\sum \limits_a q_a \rho_a^A \otimes \rho_a^B
\end{equation}
where $q_a > 0$ and $\sum \limits_{a=1}^N q_a=1$ and where 
\begin{equation}
\rho_a^A = \frac{1}{2}\bigl(\sigma_0+\vec{P}^A_a\vec{\sigma} \bigr),
\quad
\rho_a^B = \frac{1}{2}\bigl(\sigma_0+\vec{P}^B_a\vec{\sigma} \bigr)
\end{equation}
Here $\sigma_0=I$ and $\vec{\sigma}=(\sigma^1,\sigma^2,\sigma^3)$ are Pauli matrices. The most general form of Dark Matter quantum state has then a form in Fano basis
\begin{equation}
\rho_{DM}=\frac{1}{4}\sum \limits_{a=1}^N q_a \sum_{\mu,\nu=0}^3 P^A_{a,\mu}    
P^B_{a,\nu} \sigma^\mu \otimes \sigma^\nu = \frac{1}{4} \sum_{\mu,\nu=0}^3 t_{\mu \nu} \sigma^\mu \otimes \sigma^\nu 
\end{equation}


\begin{thebibliography} {}

\bibitem{einstein16} A.~Einstein, {\sl Die Grundlage der allgemeinen Relativit\"{a}tstheorie}, Annalen der Physik, {\bf 49}, 1916. Translation {\sl The Foundation of the General Theoty of Relativity} in  "A Stubbornly Persistent Illusion: The Essential Works of Albert Einstein", Edited by S. Hawking, Running Press 2007.



\bibitem{perlmutter97} S.~Perlmutter {\sl et al.}, {\sl Measurements of the Cosmological Parameters Omega and Lambda from the the First Seven Supernovae at $z \leq 0.35$}, Astrophys.J. {\bf 483}, 565 (1997).

\bibitem{perlmutter99} S.~Perlmutter {\sl et al.}, {\sl Measurements of Omega and Lambda from 42 High Redshift Supernovae}, Astrophys.J. {\bf 517}, 565 (1999).

\bibitem{riess98} A.G.~Riess {\sl et al.}, {\sl Observational Evidence for an Accelerating Universe with a Cosmological Constant}, Astronom.J. {\bf 116}, 1009 (1998).

\bibitem{schmidt98} B.P.~Schmidt {\sl et al.}, {\sl The high-z Supernova Search: Measuring Cosmic Deceleration and Global Curvature of the Universe Using Type Ia Supernov{\ae}}, Astrophys.J. {\bf 507}, 46 (1998).

\bibitem{huterer17} D.~Huterer and D.L.~Shafer, {\sl Dark energy two decades after: Observables, probes, consistency tests}, Rep.Prog.Phys. {\bf 81}, 016901 (2018).

\bibitem{planck15} Planck Collaboration, P.A.R.~Ade {\sl et al.}, {\sl Planck 2015 results, XIII. Cosmological parameters},Astronomy $\&$ Astrophysics 594, A13(2016), arXiv:1502.01589, (2015).

\bibitem{amendola10} Luca Amendola and Shinji Tsujikawa, {\sl Dark Energy - Theory and Observations}, Cambridge University Press, 2010.

\bibitem{joyce16} A.~Joyce, L.~Lombriser and F.~Schmidt, {\sl Dark Energy vs Modified Gravity}, Annu.Rev.Nuc.Part.Sci.66: 95(2016).

\bibitem{brax18} Ph.~Brax, {\sl What makes the Universe to accelerate? A review on what dark energy could be and how to test it}, Rep.Prog.Phys. {\bf 81}, 016902 (2018).

\bibitem{DES} Dark Energy Survey (DES), http://www.darkenergysurvey.org.

\bibitem{LSST} Large Synoptic Survey Telescope (LSST), http://www.lsst.org.

\bibitem{Euclid} Euclid Mission, http://sci.esa.int/euclid/.

\bibitem{WFIRST} Wide Field Infrared Survey Telescope (WFIRST), wfirst.gafc.nasa.gov.

\bibitem{spergel15} D.~Spergel {\sl et al.}, {\sl Wide-Field InfraRed Survey Telescope-Astrophysics Focused Telescope Assets}, arXiv:1503.03757 [astro-ph.CO] (2015).

\bibitem{SKA} Square Kilometer Array (SKA), arXiv:1603:01951 [astro-ph.IM] (2016). 

\bibitem{scott18} D.~Scott, {\sl The Standard Model of Cosmology: A Skeptic's Guide}, arXiv:1804.01318 [astro-ph.CO] (2018).

\bibitem{svec17b} M.~Svec, {\sl Quantum Structure of Spacetime and its  Entropy in a Cyclic Universe with Negative Curvature II: Data Analysis and Results}, arXiv:1810.06446 [gr-qc], (2018).

\bibitem{svec17a} M.~Svec, {\sl Serial Acceleration-Deceleration Transitions in a Cyclic Universe with Negative Curvature}, arXiv:1809.07178 [physics.gen-ph],  (2018).

\bibitem{bell64} J.S.~Bell, {\sl On the Einstein Podolski Rosen Paradox}, Physics {\bf 1}, 195 (1964).

\bibitem{bell87} J.S.~Bell, {\sl Speakable and unspeakable in quantum mechanics}, Cambridge University Press, 1987.

\bibitem{nielsen00} M.A.~Nielsen and I.L.~Chuang, {\sl Quantum Computation and Quantum Information}, Cambridge University Press, 2010.

\bibitem{bengtsson06} I.~Bengtsson and K.~\.{Z}yczkowski, {\sl Geometry of Quantum States - An Introduction to Quantum Entanglement}, Cambridge University Press, 2006.

\bibitem{vedral06} V.~Vedral, {\sl Introduction to Quantum Information Science}, Oxford University Press, 2006.

\bibitem{friis16} N.~Friis, S.~Bulusu and R.~Bertlmann, {\sl Geometry of two-qubit states with negative conditional entropy}, J.Phys.A:Math.Theor.{\bf 50}, 125301 (2017).

\bibitem{amari93} Shun-ichi Amari and Hiroshi Nagaoka, {\sl Methods of Information Geometry}, Translations of Mathematical Monographs, Volume 191, Oxford University Press, 1993. 

\bibitem{weinberg08} S.~Weinberg, {\sl Cosmology}, Oxford University Press, 2008.

\bibitem{carroll04} S.M.~Carroll, {\sl Spacetime and Geometry}, Addison Wesley, 2004.

\bibitem{bronshtein15} I.N.~Bronshtein, K.A.~Semendyayev, G.~Musiol and H.~M\"uhlig, {\sl  Handbook of Mathematics, Sixth Edition}, Springer Verlag, 2015.

\bibitem{polyanin02} A.D.~Polyanin and V.F.~Zaitsev, {\sl Handbook of Exact Solutions for Ordinary Differential Equations}, Second Edition, Chapman $\&$ Hall/CRC, 2002.

\bibitem{greiner94} W.~Greiner, L.~Neise and H.~St\"{o}cker, {\sl Thermodynamics and Statistical Mechanics}, Springer Verlag, 1994.

\bibitem{ryden03} B.~Ryden, {\sl Introduction to Cosmology}, Addison Wesley, 2003.

\bibitem{serjeant10} S.~Serjeant, {\sl Observational Cosmology}, Cambridge University Press, 2010.

\bibitem{lyth09} D.H.~Lyth and A.R.~Liddle, {The primordial density perturbations}, Cambridge University Press, 2009.

\bibitem{caprini18} C.~Caprini and D.G.~Figueroa, {\sl Cosmological Backgrounds of Gravitational Waves}, arXiv:1801.04268 [astro-ph.CO], (2018).

\bibitem{clarkson07} C.~Clarkson, M.~Cort\^{e}s and B.~Bassett, {\sl Dynamical Dark Energy or Simply Cosmic Curvature?}, JCAP {\bf 8}, 011 (2007).

\bibitem{PDG2015} Particle Data Group, {\sl Review of Particle Properties}, Chinese Physics {\bf C40}, Number 10, p.120 (2016).

\bibitem{PDG2014} Particle Data Group, {\sl Review of Particle Properties}, Chinese Physics {\bf C38}, Number 9, p.111 (2014).

\bibitem{jackson75} J.D.~Jackson, {\sl Classical Electrodynamics}, Second Edition, John Wiley and Sons, 1975.

\bibitem{tijms04} H.~Tijms, {\sl Understanding Probability}, Cambridge University Press, 2004.

\bibitem{landauer61} R.~Landauer, {\sl Irreversibility and heat generation in the computing process}, {\sl IBM Journal of Research and Development} {\bf 5},183 (1961).

\bibitem{peterson16} J.P.S.~Peterson {\sl et al.}, {\sl Experimental demonstration of information to energy conversion in a quantum system at the Landauer Limit}, Proc.R.Soc. {\bf 472}, 20150813 (2016). 

\bibitem{yan18}  L.L.~Yan {\sl et al.}, {\sl  Single-atom demonstration of quantum Landauer principle}, arXiv:1803.10424 [quant-ph] (2018).

\bibitem{prigogine77} G.~Nicolis and I.~Prigogine, {\sl Self-organization in Non-equilibrium Systems: From Dissipative Structures to Order through Fluctuations}, Wiley-Interscience, 1977.

\bibitem{jantsch80} E.~Jantsch, {\sl The Self-Organizing Universe - Scientific and Human Implications of the Emerging Paradigm of Evolution}, Pergamon Press, 1980.

\bibitem{kauffman93} S.~Kauffman, {\sl The Origins of Order: self-organization and selection in evolution}, Oxford University Press, 1993. 

\bibitem{pathria72} R.K.~Pathria, {\sl  Statistical Mechanics}, Pergamon Press, 1972, p.147. 

\bibitem{tolman34}  R.C.~Tolman, {\sl Relativity, Thermodynamics and Cosmology}, Oxford University Press, 1934.

\bibitem{svec17c} M.~Svec, {\sl From emergent gravity to galactic rotation curves}, 2018, to appear.
 
\bibitem{mcgaugh16} S.~McGaugh, F.~Lelli and J.M.~Schombert, {\sl The Radial Acceleration Relation in Rotation Supported Galaxies}, Phys.Rev.Lett. {\bf 117}, 201101 (2016).

\bibitem{lelli16} F.~Lelli, S.S.~McGaugh, J.M.~Schombert and M.S.~Pawlowski, {\sl One Law to Rule Them All: The Radial Acceleration Relation of Galaxies}, Astrophys.J. 830, 152(2017). 

\bibitem{hawking82} S.H.~Hawking, {\sl The Unpredictability of Quantum Gravity}, Commun.Math.Phys. {\bf 87}, 395 (1982).

\bibitem{hawking84} S.H.~Hawking, {\sl Non-Trivial Topologies in Quantum Gravity}, Nucl.Phys. {\bf B244}, 135 (1984).

\bibitem{svec13a} M.~Svec, {\sl Study of $\pi N \to \pi \pi N$ processes on polarized targets: Quantum environment and its dephasing interaction with particle scattering}, Phys.Rev. {\bf D91}, 074005 (2015).


\end{thebibliography}
\end{document}